\documentclass[preprint,preprintnumbers,amsmath,amssymb]{revtex4}
\usepackage{graphicx}% Include figure files
\usepackage{dcolumn}% Align table columns on decimal point
\usepackage{bm}% bold math
\usepackage{verbatim}
\usepackage{xcolor}
\usepackage{dsfont}
\usepackage{epsfig}
\usepackage{bbm}
\usepackage{epstopdf}
\usepackage{hyperref}
\usepackage{amsmath,mathtools}
\usepackage{natbib}
\usepackage[caption=false]{subfig}
\usepackage{dsfont}
\usepackage{multirow}
\usepackage{physics}
\usepackage{cellspace}
\newcommand\beq{\begin{equation}}
\newcommand\eeq{\end{equation}}
\newcommand\bea{\begin{eqnarray}}
\newcommand\eea{\end{eqnarray}}

\def\sgn{{\rm sgn}}
\usepackage{xcolor}

\def\Tr{{\rm Tr}}

\def\id{\mathds{1}}

\newcommand\LG[1]{\rm LG{#1}}

\def\x0{{{\bf x}_0}}

\begin{document}

	%\preprint{Imperial/TP/***}
	
	\title{Leggett-Garg violations for continuous variable systems with gaussian states}

	\author{C. Mawby}
	\email{c.mawby18@imperial.ac.uk}
	\author{J.J.Halliwell}%
	\email{j.halliwell@imperial.ac.uk}
	\numberwithin{equation}{section}
	\renewcommand\thesection{\arabic{section}}

	\affiliation{Blackett Laboratory \\ Imperial College \\ London SW7
		2BZ \\ UK }

	
	\begin{abstract}
	Macrorealism (MR) is the worldview that certain quantities may take definite values at all times irrespective of past or future measurements and may be experimentally falsified via the Leggett-Garg (LG) inequalities.  We put this world view to the test for systems described by a continuous variable $x$ by seeking LG violations for measurements of a dichotomic variable $Q = \textrm{sign}(x)$, in the case of gaussian initial states in a quantum harmonic oscillator. Extending our earlier analysis [C. Mawby and J. J. Halliwell, Phys. Rev. A \textbf{105}, 022221 (2022)] we find analytic expressions for the temporal correlators. An exploration of parameter space reveals significant regimes in which the two-time LG inequalities are violated, and likewise at three and four times.  To obtain a physical picture of the LG violations, we exploit the continuous nature of the underlying position variable and analyze the relevant quantum-mechanical currents, Bohm trajectories, and Wigner function.  We also show that larger violations are possible using the Wigner LG inequalities.  Further, we extend the analysis to LG tests using coherent state projectors, thermal coherent states, and squeezed states.
     \end{abstract}
	\maketitle

\section{Introduction}

The motion of a pendulum has been used by clockmakers and hypnotists alike for centuries, with its regular left, right, left motion.  Scaled down enough, the quantum mechanical description becomes necessary, which hints that non-classical states without a definite left-right property underlie the motion.  The existence of these type of states form one of the pillars of many quantum technologies, and hence verification of their existence, and an understanding of their persistence to macroscopic scales is of great interest.

The Leggett-Garg inequalities~\cite{leggett1985, leggett1988,leggett2008,emary2013} were introduced to provide a quantitative test capable of demonstrating the failure of the precise world view known as macrorealism (MR).  MR is defined as the conjunction of three realist tenets -- that a system resides in one observable state only, for all instants of time, which may be measured without influencing future dynamics of the system, and that measurements respect causality.  The violation of these inequalities indicates a failure of MR, and hence the presence of non-classical behaviour.

The LG inequalities are typically established for a dichotomic observable $Q$, which may take value $s_i=\pm 1$, measured in a series of experiments at single times and at pairs of times.
This yields a data set consisting of single time averages $\ev{Q_i}$, where $Q_i = Q(t_i)$, and the temporal correlators $C_{ij}$ defined by
\begin{equation}
C_{ij}=\ev{Q_i Q_j} = \sum_{i,j}s_i s_j p(s_i, s_j),
\end{equation}
where $p(s_i, s_j)$ is the two-time measurement probability, giving the likelihood of measuring $s_i$ and $s_j$ at times $t_i$, $t_j$.
The temporal correlators must be measured in a non-invasive manner, in keeping with the definition of MR, which is typically done using ideal negative measurements~\cite{knee2012,robens2015,katiyar2017} (but other approaches exist~\cite{halliwell2016a,majidy2019a,majidy2019b,zaw2022a,tsirelson2006}).

For the commonly-studied three time case, the data set consists of three correlators $C_{12}, C_{23}, C_{13}$ and three single time averages $\ev{Q_i}$, $ i=1,2,3$.
These quantities then form six puzzle-pieces that, should the system obey the assumptions of MR, must be the moments of an underlying joint probability distribution $p(s_1, s_2, s_3)$.  In the case where this endeavour is possible, the correlators and averages will satisfy the three-time LG inequalities (LG3):
\bea
L_1=1 + C_{12} + C_{23} + C_{13} & \ge & 0,
\label{LG1}
\\
L_2=1 - C_{12} - C_{23} + C_{13} & \ge & 0,
\label{LG2}
\\
L_3=1 + C_{12} - C_{23} - C_{13} & \ge & 0,
\label{LG3}
\\
L_4=1 - C_{12} + C_{23} - C_{13} & \ge & 0.
\label{LG4}
\eea
They will also satisfy a set of twelve two-time LG inequalities (LG2), four of which are of the form,
\beq
1+s_1\ev{Q_1}+s_2 \ev{Q_2}+s_1 s_2 C_{12}\geq 0
\label{LG22}
\eeq
where $ s_1, s_2 = \pm 1 $, with two more sets of four for the other two time pairs.
This set of sixteen inequalities are necessary and sufficient conditions for MR
\cite{halliwell2016b,halliwell2017,halliwell2019b,halliwell2019, halliwell2020,araujo2013,majidy2021b}.
If any one of them is violated, MR fails. From an experimental point of view, the LG2 inequalities constitute the simplest place to look first, since only one correlator needs to be measured.

For measurements at four times, there are six correlators, but the natural data set is a cycle of four $C_{12}, C_{23}, C_{34}, C_{14} $, along with the four single time averages.
The $\LG{4}$ inequalities take the form 
\begin{equation}
\label{eq:LG4}
-2\leq C_{12}+C_{23}+C_{34}-C_{14}\leq 2,
\end{equation}
together with the six more inequalities permuting the location of the minus sign. The necessary and sufficient conditions for MR at four times consist of these eight LG4 inequalities, together with the set of sixteen LG2s for the four time pairs \cite{halliwell2016b, halliwell2019, halliwell2017,halliwell2019b}.

This general framework has been put to the test experimentally in many types of systems. See for example Refs. \cite{majidy2019a,goggin2011,formaggio2016, joarder2022,knee2012, robens2015,katiyar2017}  and also the useful review \cite{emary2013} . Most of these experiments are on systems that are essentially microscopic but some come close to macroscopicity \cite{knee2016}. Note also the useful critique of the LG approach Ref.~\cite{maroney2014a}.

In the present paper we investigate LG tests for the quantum harmonic oscillator (QHO). The LG framework is readily adapted to this physical situation using a single dichotomic variable $Q = {\rm sgn} (x)$, where $x$ is the particle position, measured at different times. We build on our earlier work Ref. \cite{mawby2022}, which explored LG violations in the QHO for initial states consisting of harmonic oscillator eigenstates and superpositions thereof, finding close to maximal violations in some cases. Here we focus on the case of an initial coherent state and closely related states. Such an initial state is intriguing in this context since it is often regarded as essentially classical, being a phase space localized state evolving along a classical trajectory. Hence any LG violations arising from this state constitute particularly striking examples of non-classical behaviour.

LG tests for the QHO with an initial coherent state were first explored by Bose et al \cite{bose2018}, who proposed an experiment to measure the temporal correlators and carried out calculations of LG4 violations. A subsequent paper \cite{das2022a} explored both LG2 violations and also violations of the no-signalling in time conditions \cite{kofler2013,clemente2015,clemente2016}. Our work in part parallels Ref. \cite{das2022a}.

The first main aim of this paper is to undertake a thorough analysis of LG2, LG3 and LG4 inequalities for an initial coherent state. This is carried out in Section \ref{sec:corecalc}, where we set out the formalism, and drawing on our earlier work calculate the temporal correlators \cite{mawby2022}.  We carry out a detailed parameter search and find the largest LG violations possible for a coherent state.

A second aim is to explore the physical origins of the LG violations. So, in Section \ref{sec:physmec} we examine the difference between quantum-mechanical currents and their classical counterparts for initial coherent states projected onto the positive or negative $x$-axis.  In this section we also provide a second approach to calculating correlators in the small-time limit, which is in fact valid for general states.  We also calculate the Bohmian trajectories, to give further physical portrait of what underlies the observed LG violations.  Finally, we examine the measurement process in the Wigner representation, noting that the initially positive Wigner function of the coherent state acquires negativity as a result of the projective measurement process. 

Modifications to the above framework are considered in Section \ref{sec:mf}. We investigate what violations are possible using the Wigner LG inequalities \cite{saha2015, naikoo2020}.  We briefly discuss other types of measurements beyond the simple projective position measurements used so far and also consider LG2 violations with projections onto coherent states. We also determine how the LG violations may be modified for squeezed states or thermal states. We summarize in Section \ref{sec:sum}.  We relegate to a series of appendices the grisly details of the calculations involved in this analysis.

\section{Calculation of Correlators and LG Violations}
\label{sec:corecalc}
\subsection{Conventions and Strategy}
For most of this paper, we will work with coherent states of the harmonic potential,  which can of course be thought of as the ground state of the QHO, shifted in phase-space.  The intricacy of calculating temporal correlators within QM stems from the complexity in the time evolution of a post-measurement state.  By considering a co-moving frame for the post-measurement state, we develop a time evolution result which explicitly separates the quantum behaviour from the classical trajectories.

We will work with systems defined exactly (or approximately) by the harmonic oscillator Hamiltonian,
\begin{equation}
\hat H = \frac{\hat p_{\text{phys}}^2}{2m}+\frac12 m\omega^2 \hat x_{\text{phys}}^2,
\end{equation}
with physical position and momenta $x_{\text{phys}}$ and $p_{\text{phys}}$.  In calculations we use the standard dimensionless variables $x\sqrt{\hbar/(m\omega)}=x_{\text{phys}}$ and $p\sqrt{\hbar m\omega}=p_{\text{phys}}$. We denote energy eigenstates $\ket n$, writing $\psi_n(x)$ in the position basis, with corresponding energies $E_n=\hbar \omega(n+\frac12)=\varepsilon_n \hbar \omega$.

We write coherent states as $\ket \alpha$, where its eigenvalue $\alpha$ relates to rescaled variables as
\begin{align}
\label{eq:alphatox}
\ev{\hat x(t)}&=\sqrt2 \Re \alpha(t),\\
\ev{\hat p(t)}&=\sqrt2 \Im \alpha(t).
\end{align}
These are the classical paths underlying the motion of coherent states, and we adopt the short-hand $x_1=\ev{\hat x(t_1)}$ and likewise $p_1=\ev{\hat p(t_1)}$.  A coherent state may be represented in terms of $\alpha$ and an initial phase, however in this work we will largely represent them in terms of the initial averages $x_0$ and $p_0$, for clarity of physical understanding.  We construct coherent states with the unitary displacement operator $D(\alpha)=\exp(\alpha a^\dagger-\alpha^* \hat a)$ operating on the ground state,
\begin{equation}
\ket \alpha = D(\alpha)\ket 0,
\end{equation}
which results in the wave-function
\begin{equation}
\psi^{\alpha}(x,t)=\frac{1}{\pi^\frac14}\exp(-\frac12(x-x_t)^2 + i \frac{p_t}{\hbar} x+i\gamma(t)).
\end{equation}
We will calculate the quantities $\ev{Q_i}$, $C_{ij}$ appearing in the LG inequalities. A convenient way to proceed is to first note that the combination appearing in the LG2 inequalities is proportional to the quantity
\begin{equation}
\label{eq:momexp}
q(s_1, s_2)=\frac{1}{4}\left(1+s_1 \langle\hat Q_1\rangle+s_2 \langle\hat Q_2\rangle+s_1 s_2 C_{12}\right).
\end{equation}
Classically this quantity is non-negative and is the probability distribution matching the data set with moments of $\ev{Q_1}$, $\ev{Q_2}$, $C_{12}$. In the quantum-mechanical case, this quantity may be written
\begin{equation}
\label{eq:qp}
q(s_1, s_2)=\text{Re} \Tr (P_{s_2}(t_2)P_{s_1}(t_1)\rho),
\end{equation}
where $s=\pm1$, and $P_{s}$ are projection operators corresponding to the measurement made, with
\begin{equation}
\label{eq:ptoq}
P_s=\frac12(1+s\hat Q).	
\end{equation}
Since Eq.~(\ref{eq:qp}) can be negative (up to a maximum of $-\tfrac18$ the L{\"u}ders bound), it is referred to as a quasi-probability (QP)~\cite{halliwell2016b,halliwell2019c}. Purely from a calculational point of view it is a convenient object to work with, as we found in Ref.~\cite{mawby2022}, since it is proportional to the LG2 inequalities in the quantum case, and since the correlators are easily read off from its moment expansion, so we make use of it here.

The QP may be rewritten in the form given in Ref.~\cite{halliwell2019c}
\begin{equation}
 q(s_1, s_2)=\frac18\ev*{(1+s_1 \hat Q_1 +s_2 \hat Q_2)^2 -1}
 \end{equation}
which makes explicit that the maximally violating state satisfies the eigenvalue equation 
\begin{equation}
\label{eq:maxQP}
(s_1\hat Q_1 + s_2 \hat Q_2)\ket\psi=-\ket\psi. 
\end{equation}
We have not been able to find the maximally violating state but Eq. (\ref{eq:maxQP}) suggests it is probably discontinuous at $x=0$. This in turn suggests that we will not able to get close to maximal violations with the simple gaussian states explored here.
\subsection{Calculation of the Correlators}
We now calculate the temporal correlators for the case $\hat Q = \sgn(\hat x)$.  The QP Eq.~(\ref{eq:qp}) is given by
%The quasi-probability for coherent states with the left-right dichotomic variable, $P_s=\theta(s\hat x)$, is given by

\begin{equation}
q(s_1,s_2)=\Re\ev{e^{\frac{i H t_2}{\hbar}}\theta(s_2\hat x)e^{-\frac{i H \tau}{\hbar}}\theta(s_1\hat x)e^{-\frac{i H t_1}{\hbar}}}{\alpha}.
\end{equation}
By considering the displacement operator as acting on the measurements instead of the state, the quasi-probability is shown in Appendix \ref{app:timeev} to be
\begin{equation}
\label{eqn:quasgen}
q(s_1,s_2)=\Re e^{\frac{i\omega \tau}{2}} \ev{\theta(s_2(\hat x + x_2))e^{-iH\tau}\theta(s_1(\hat x + x_1))}{0},
\end{equation}
which reveals that the quasi-probability for coherent states can be understood as the quasi-probability for the pure ground state, with measurement profiles translated according to the classical paths.  This shows that any LG test on any coherent state may be directly mapped to an LG test on the ground-state of the QHO, with translated measurements. 

Using the surprising result that $\int_a^b \psi_n(x)\psi_m(x) dx$ has an exact and general solution, where $\psi_n(x)$ are energy eigenstates \cite{moriconi2007a, mawby2022}, we are able to calculate the temporal correlators as an infinite sum
\begin{equation}
\label{eqn:corr}
C_{12}=\erf(x_1)\erf(x_2)+ 4\sum_{n=1}^\infty \cos (n\omega \tau) J_{0n}(x_1,\infty)J_{0n}(x_2,\infty),
\end{equation}
where
\begin{equation}
	J_{0n}(x, \infty)=\mel{0}{\theta(\hat x - x)}{n}.
\end{equation}
The $J_{0n}$ terms are given in terms of the $\psi_n(x)$ and its derivative.  The details of this calculation are given in Appendix ~\ref{app:timeev}.  The infinite sum may be evaluated approximately using numerical methods, by summing up to a finite $n$.  This calculation matches the analytically calculated special case of $x_0=0$ given in Ref. \cite{halliwell2021}.
%We now observe that the $J_{0n}$ terms in the infinite sum will be suppressed if either $\lvert x_1 \rvert$ or $\lvert x_2 \rvert$ become large, since each term involves a factor of either $\psi_0(x)$ or $\psi'_0(x)$, resulting in double-exponential suppression.  We can place a limit on the magnitude of the sum in Eq.(\ref{eqn:cohquas}), by considering it as the overlap between $\theta(x-x_1)\psi_0(x)$ and $\theta(x-x_2)\psi_0(x)$, less $J_{00}(x_1, \infty)J_{00}(x_2, \infty)$.  This yields the limit
%\begin{equation}
%\abs{\sum_{n=1}^\infty e^{-in\omega \tau}J_{0n}(x_1,\infty)J_{0n}(x_2,\infty)}\leq\frac{1}{2} \theta (x_1-x_2) (\text{erf}(x_2)+\text{erfc}(x_1)-1)-\frac{1}{4} (\text{erfc}(x_1)-2) \text{erfc}(x_2).
%\end{equation}
%A quick evaluation shows that with either $|x_1|$ or $|x_2|$ greater than $3$, the magnitude of the sum is limited to the order of $10^{-6}$.  This means that if the coherent state is a significant distance from the axis at either of the measurement times, the sum containing the quantum effects will vanish, and the quasi-probability will be non-negative.
%
%The same applies to the correlators Eq.~(\ref{eqn:corr}) used in the LG inequalities, where any $x_i>3$ renders the correlators involving that measurement time classical.  This does not rule out the possibility of LG3/4 violations with $x_i>3$, but it does mean at least two of the correlators will be forced to be classical, which will clearly reduce in magnitude any possible violations.
\subsection{LG Violations}
\label{sec:LGresults}
\begin{figure}
	\subfloat[]{{\includegraphics[height=5.6cm]{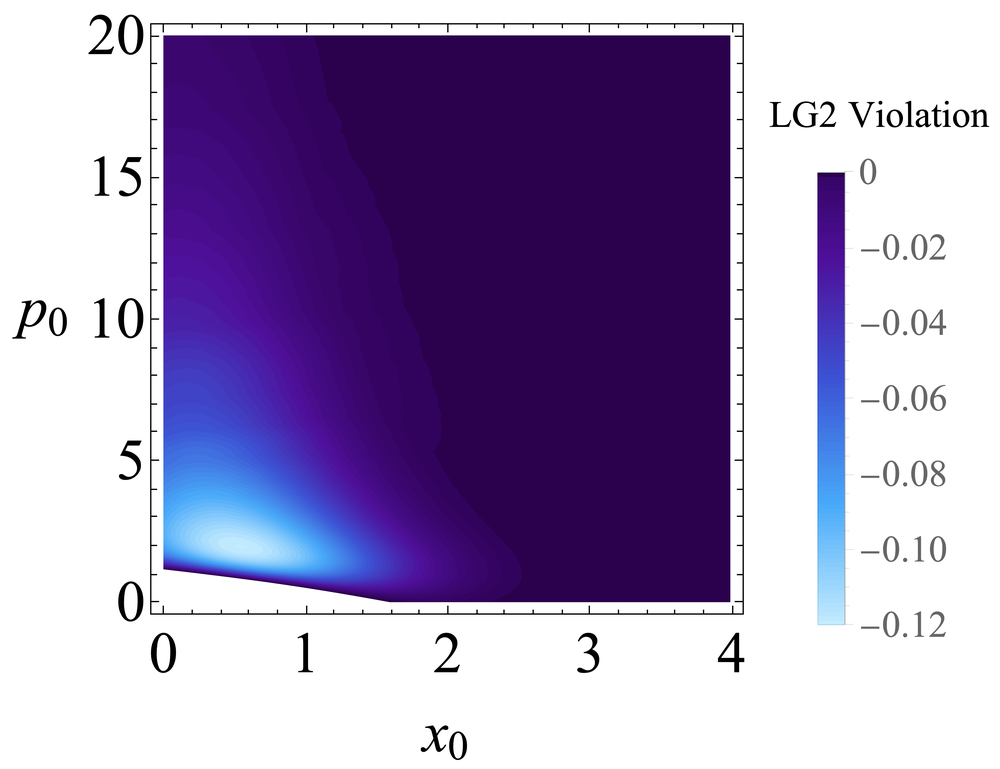}}}%
	%\qquad
	\hspace{5mm}
	\subfloat[]{{\includegraphics[height=5.6cm]{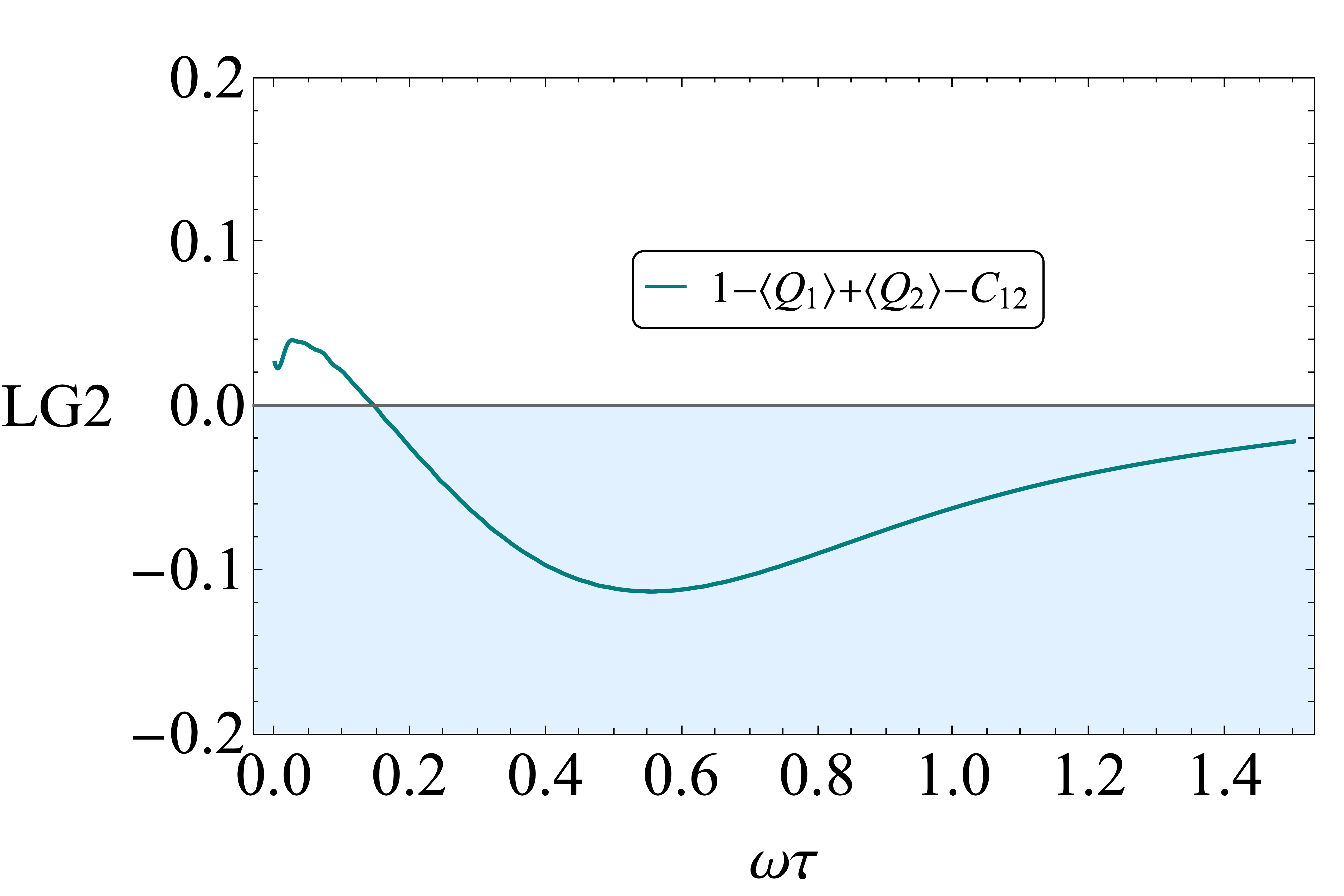}}}
	\caption{Plot (a) is a parameter space exploration, showing the largest LG2 violation for a given coherent state. Plot (b) shows the temporal behaviour of the LG2s for a state leading to the largest violation of $-0.113$.}%
	\label{fig:lg2}%
\end{figure}
The freely chooseable parameters are the initial parameters of the coherent state $x_0$, $p_0$, and the time interval between measurements, $\tau$.  Where there are more than two measurements, we use equal time-spacing. In Fig.~\ref{fig:lg2} (a), we plot a parameter space exploration of violations for the LG2 inequalities, where for a given $x_0$, $p_0$, we have numerically searched for the largest violation for that state. The LG3 and LG4 inequalities have a similar distribution, but with progressive broadening. Figures for the LG3 and LG4 inequalities are included in Appendix~\ref{app:LGresults}.

\begin{figure}
	\subfloat[]{{\includegraphics[height=5.2cm]{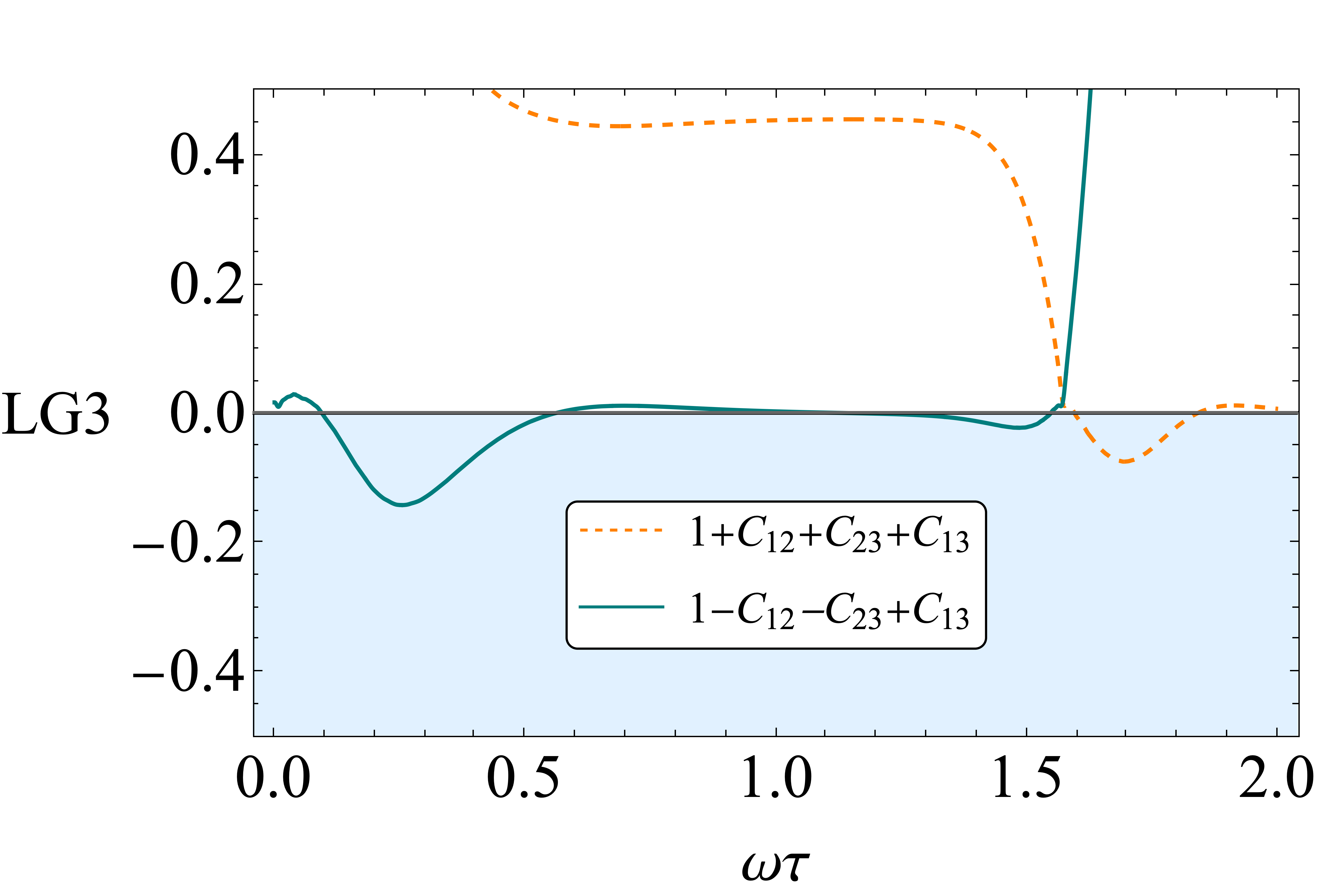}}}%
	%\qquad
	\hspace{5mm}
	\subfloat[]{{\includegraphics[height=5.2cm]{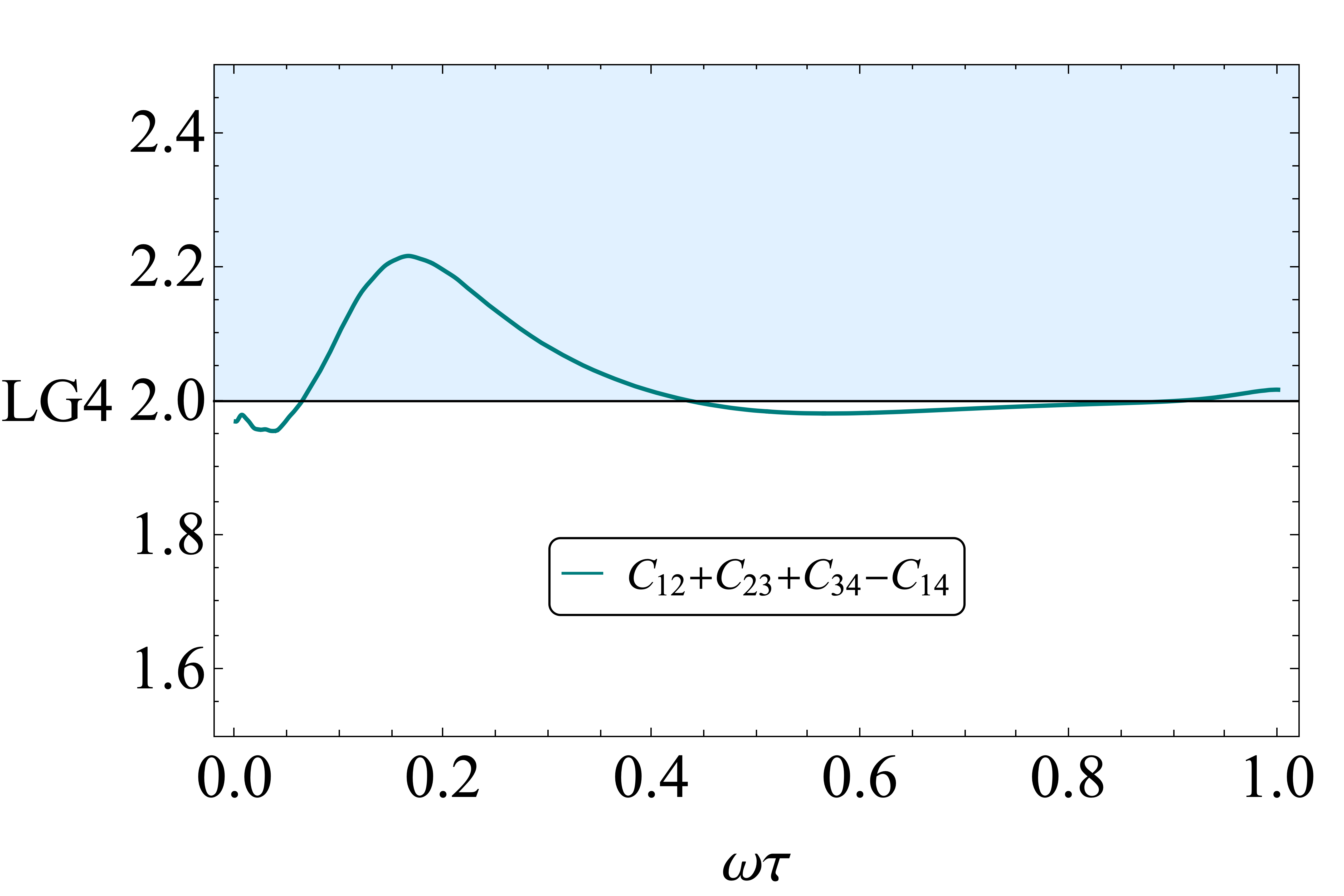}}}
	\caption{Plot (a) shows temporal behaviour of the LG3s for a coherent state with the largest violation $-0.141$. Plot (b) shows the same for the LG4s with a coherent state leading to the largest violation of $2.216$.}%
	\label{fig:lg34}%
\end{figure}

The parameters leading to the largest violation, and the magnitude of those violations are reported in Table~\ref{tab:1}.  In Fig.~\ref{fig:lg2}(b) and Fig.~\ref{fig:lg34}, we plot the temporal behaviour of the LG2, LG3 and LG4 violations for the states in Table~\ref{tab:1} leading to the largest violations.  

\begin{table}
\centering
\setlength{\tabcolsep}{0.5em}
\begin{tabular}{|c|c|c|c|}
\hline
Inequality         & Largest Violation & Percent of L{\"u}ders Bound & Location ($|x_0|, |p_0|$) \\ \hline
LG2 & -0.113            & 22\%                    & (0.550, 1.925)        \\
LG3                        & -0.141            & 28\%                    & (0.859, 3.317)           \\
LG4                        & 2.216              & 26\%                    & (0.929, 3.666)          
\\
\hline
\end{tabular}
\caption{\label{tab:1}Tabulation of parameter space results.}
\end{table}

\section{Physical Mechanisms of Violation}
\label{sec:physmec}
Given the LG violations exhibited in the previous section a natural question to ask concerns the underlying physical effects producing the non-classical behaviour  responsible for the violations. Since $ Q(t) = \sgn(x(t) )$, a classical picture of the system would involve a set of trajectories $x(t)$ and probabilities for those trajectories. It is then natural to look at the parallel structures in quantum theory and compare with the classical analogues. We therefore look at the quantum-mechanical currents associated with the LG inequalities, which correspond to the time evolution of certain probabilities, and also to the Bohm trajectories associated with those currents, in terms of which the probability flow in space-time is easily seen.
What we will see is that the departures from classicality are essentially the ``diffraction in time’’ effect first investigated by Moshinsky, who considered the time evolution of an initial plane wave  in one dimension restricted to $x<0$ \cite{moshinsky1952, moshinsky1976}.  The key mathematical object is the Moshinsky function 
\begin{equation}
\label{eq:moshfunc}
M(x,p,t)=\langle x\lvert e^{-iHt}\theta(\hat x)\rvert p \rangle 
\end{equation}
for an initial momentum state $\lvert p\rangle$, which we will see below appears in the calculation of the quasi-probability.

\subsection{Analysis with Currents}

As we saw in Section.~\ref{sec:LGresults}, the quasi-probability component $q(-,+)$ exhibits a healthy degree of negativity.  We start by writing it as
\begin{equation}
\label{eq:dqdt}
q(-,+)=\int_{0}^{\tau}\mathop{dt}\frac{dq}{dt}.
\end{equation}
 It is then simple to relate $\frac{dq}{dt}$ to a set of quantum mechanical currents, which can be calculated analytically.  Overall negativity of the QP can then be spotted by the non-classicality or negativity of certain combinations of currents.  With details in Appendix \ref{app:curr}, we are able to write the quasi-probability as the following combinations of currents at the origin,
\begin{equation}
\label{eq:qpdiffs}
q(-,+)=\int_{t_1}^{t_2}\mathop{dt}J_-(t)+\frac12 \int_{t_1}^{t_2}\mathop{dt}\left(\mathbb{J}_-(t)-J_-(t) +\mathbb{J}_+(t)-J_+(t)\right),
\end{equation}
where $J_\pm(t)$ is the current following a measurement of $s_1 = \pm$, and $\mathbb{J}_\pm(t)$ are the classical analogues of this.  The chopped currents contains the complexity of the influence of the earlier measurement, and are hence quite complicated and given in Appendix~\ref{app:chopcurr}.
 
Note that the first time integral is simply the sequential measurement probability $p_{12}(-,+)$, which is non-negative.  The negativity of the quasi-probability therefore arises as a result of the difference between the classical and quantum chopped currents. 

\begin{figure}
	\subfloat[]{{\includegraphics[height=5.2cm]{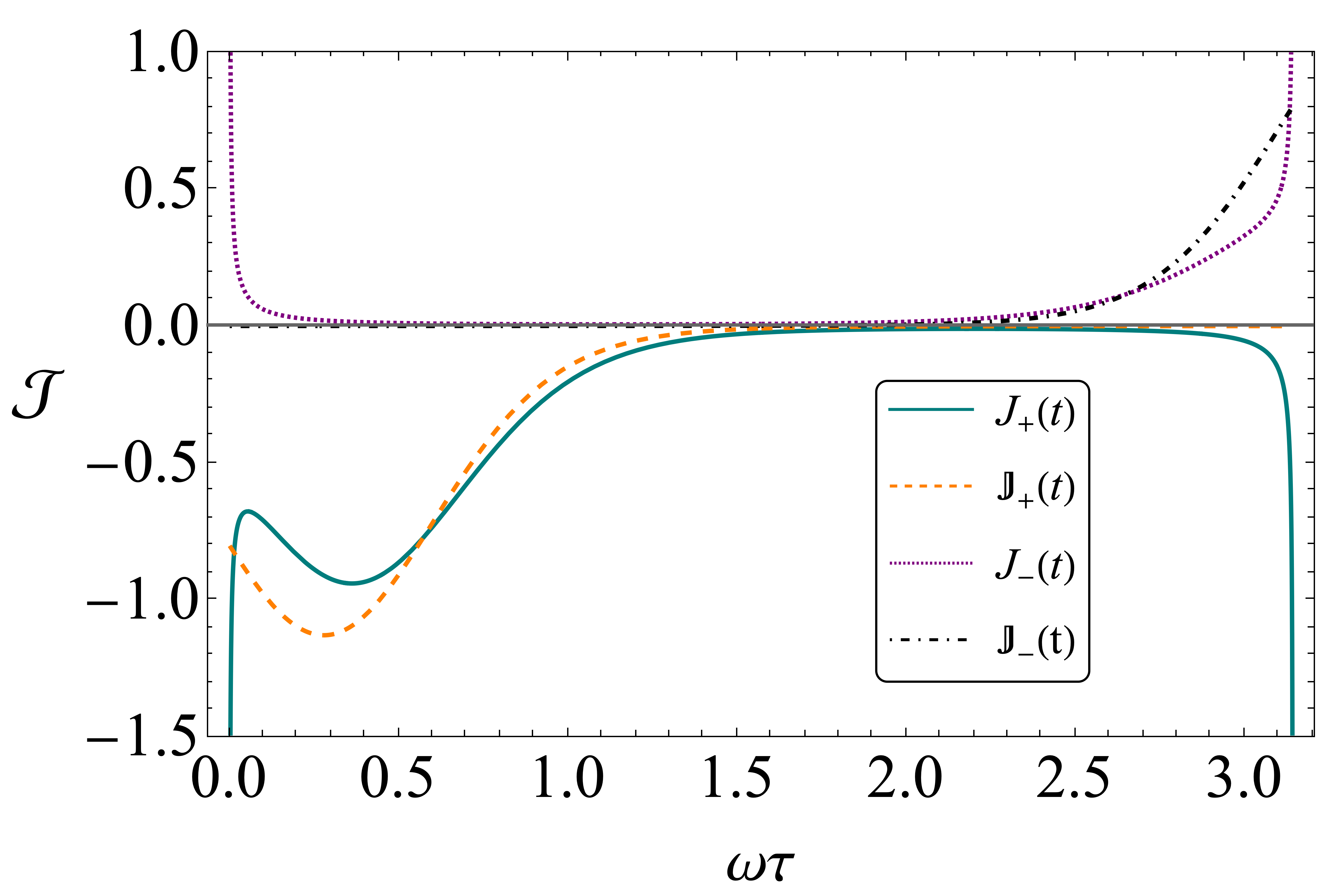}}}%
	%\qquad
	\hspace{5mm}
	\subfloat[]{{\includegraphics[height=5.2cm]{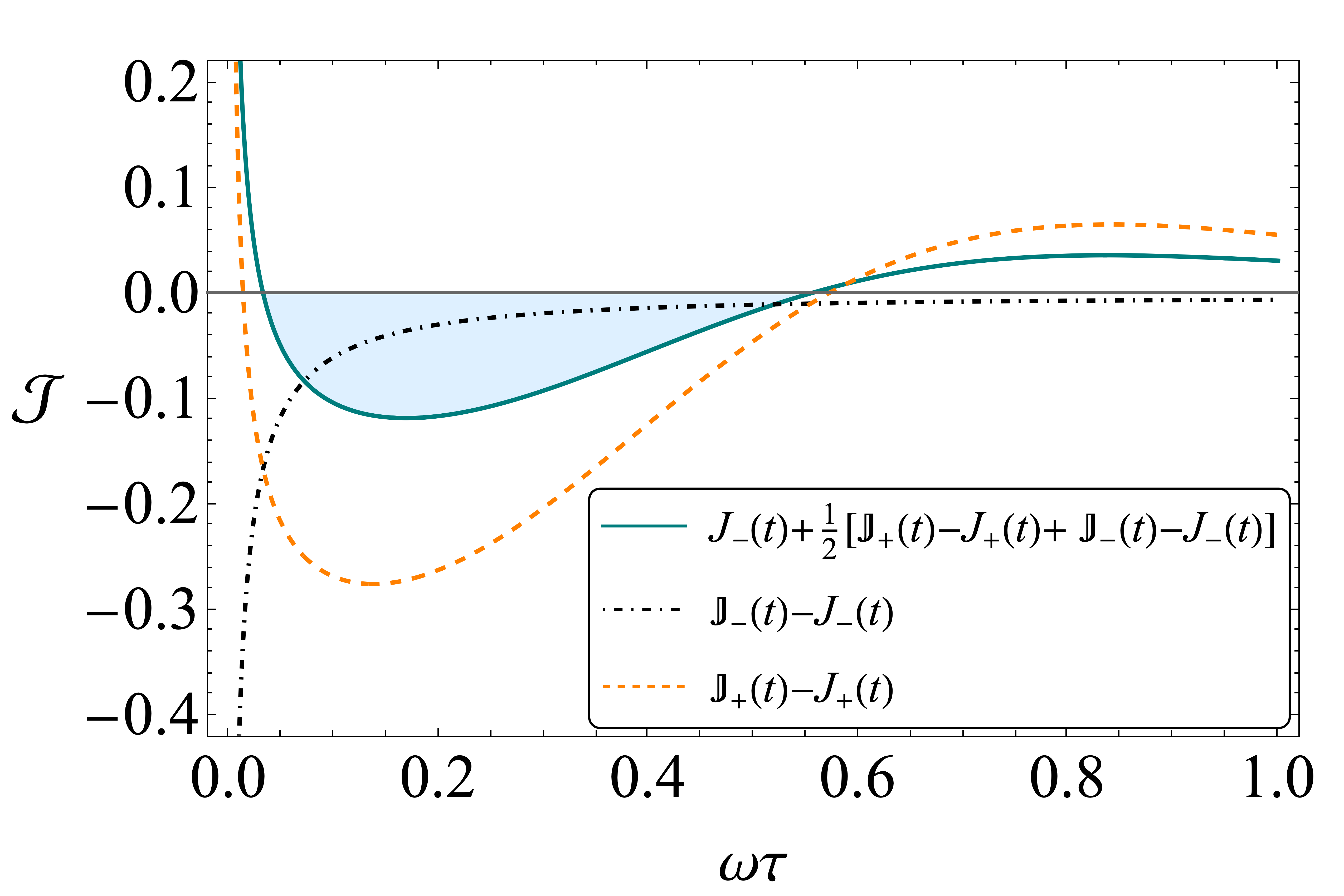}}}
	\caption{Plot (a) shows the post-measurement quantum currents $J_\pm(t)$, and their classical analogues $\mathbb{J}_\pm(t)$, normalised to $\mathcal{J}=\tfrac{J}{\omega}$, for the coherent state with $x_0=0.55$, $p_0=-1.925$. Plot (b) shows their combination  appearing in the time derivative of $q(-,+)$, Eq.~(\ref{eq:qpdiffs}).}%
	\label{fig:currs}%
\end{figure}

The classical and quantum chopped currents and the current combination appearing in Eq.~(\ref{eq:qpJ}) are all plotted in Figure~\ref{fig:currs} for the initial state giving the LG2 violation described in Section \ref{sec:LGresults}. The departures from classicality are clearly seen and are consistent with a broadening of the momentum distribution produced by the measurement. Note also that the quantum chopped currents diverge initially due to the sharpness of the measurement.

Most importantly, we see that the combination of currents appearing in the quasi-probability Eq.~(\ref{eq:qpJ}) will clearly produce an overall negativity when integrated over time, thereby confirming the LG2 violation shown in Fig.~\ref{fig:lg2}. Furthermore, we have integrated Eq.~(\ref{eq:qpJ}) numerically and find an exact agreement with the calculation of Section \ref{sec:corecalc}, Eq.~(\ref{eqn:cohquas}) thereby providing an independent check of this result. 

It is also convenient to explore the currents in the small time limit. This is done in App \ref{app:tJ} and gives a clear analytic picture of the departures from classicality. Since these expressions are valid for any initial state they could provide a useful starting point in the search for other initial states giving  LG2 violations larger than the somewhat modest violations found here.

\subsection{Bohm trajectories}
\begin{figure}
	\includegraphics[height=9.4cm]{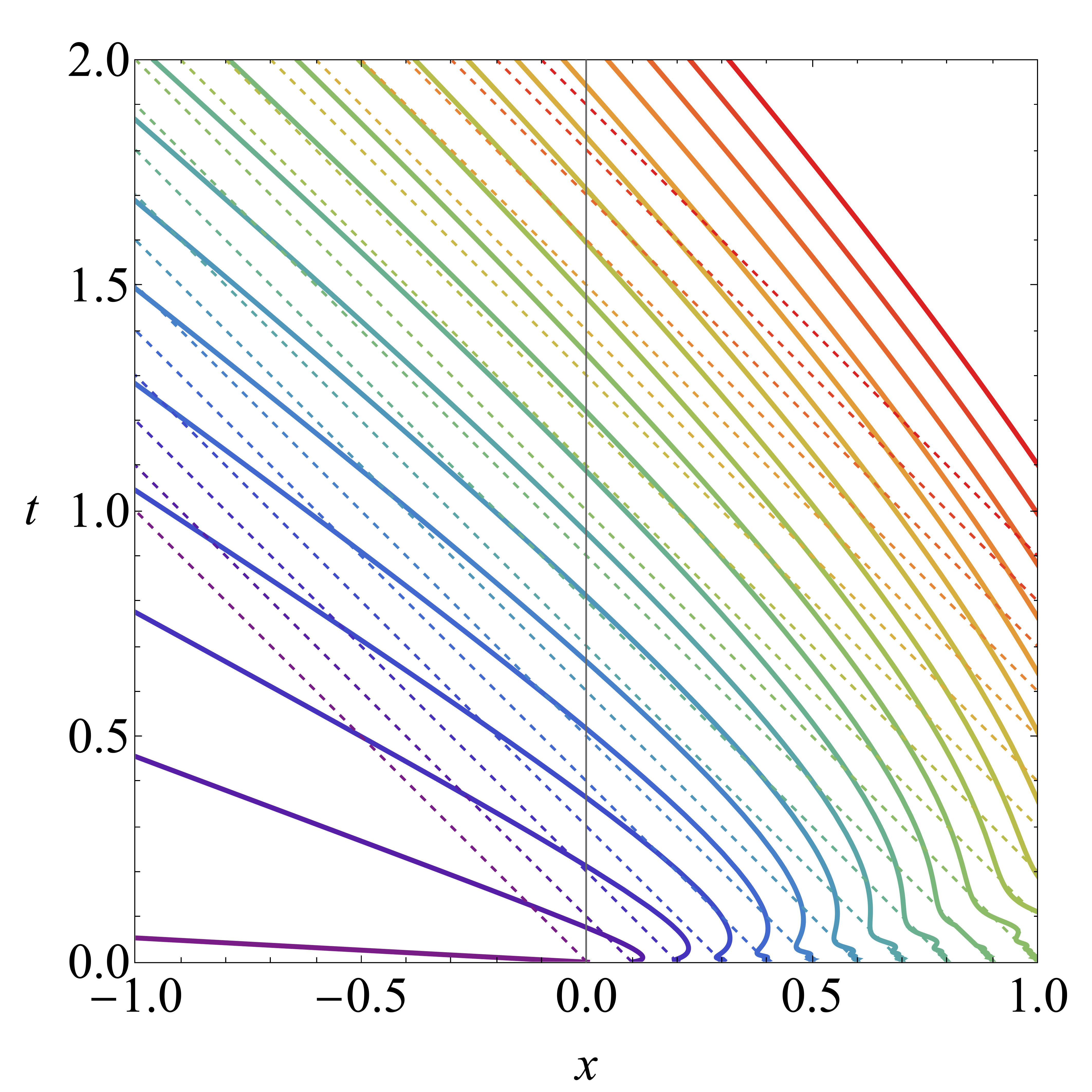}
	\caption{The Bohm trajectories associated with the Moshinsky function, $\langle x|e^{-iHt}\theta(\hat x)|p\rangle$ with $p = -1$.  The equivalent classical paths are shown dotted.}%%
	\label{fig:bohmMosh}
\end{figure}
To give a visual demonstration of how the measurements influence motion in a way to lead to LG2 violations, we now calculate and plot the de Broglie-Bohm trajectories~\cite{bohm1952,bohm1952a,debroglie1927}.

As noted earlier, the Moshinsky function underlies the behaviour of the quasi-probability for these measurements, so we initially examine the Bohm trajectories for this scenario.  Using Moshinsky's calculation (free particle dynamics), we calculate the quantum-mechanical current $J_M(x,t)$, which we then use in the guidance equation for Bohm trajectories,
\begin{equation}
\dot{x}(t)=\frac{J_M(x,t)}{\lvert M(x,p,t)\rvert^2},
\end{equation}
which we proceed to solve numerically.

In Fig.~\ref{fig:bohmMosh}, we plot the trajectories for a state initially constrained to the right-hand side of the axis, with a left-ward momentum, with classical trajectories shown dotted.  

From this we see two distinct phases of deviation from the classical result.  Initially the trajectories rapidly exit the right hand side, with a negative momentum larger than in the classical case, an anti-Zeno effect \cite{kaulakys1997}.  After a short while, a Zeno effect \cite{turing2001, misra1977} happens, and the trajectories bend back relative to the classical trajectories, staying in the right hand side longer than in the classical case.  We will see both of these behaviours at play in the case studied in this paper.

Using the expressions for the chopped current Eq.~(\ref{eq:chopJ}) and chopped wave-function Eq.~(\ref{eq:chopwav}), we can write the guidance equation for the harmonic oscillator case,
\begin{equation}
	\dot x (t) = \frac{J_{\pm}(x,t)}{\lvert \phi_\alpha^{\pm}(x,t)\rvert^2},
\end{equation}
which we again solve numerically.

\begin{figure}
	\subfloat[]{{\includegraphics[height=7.2cm]{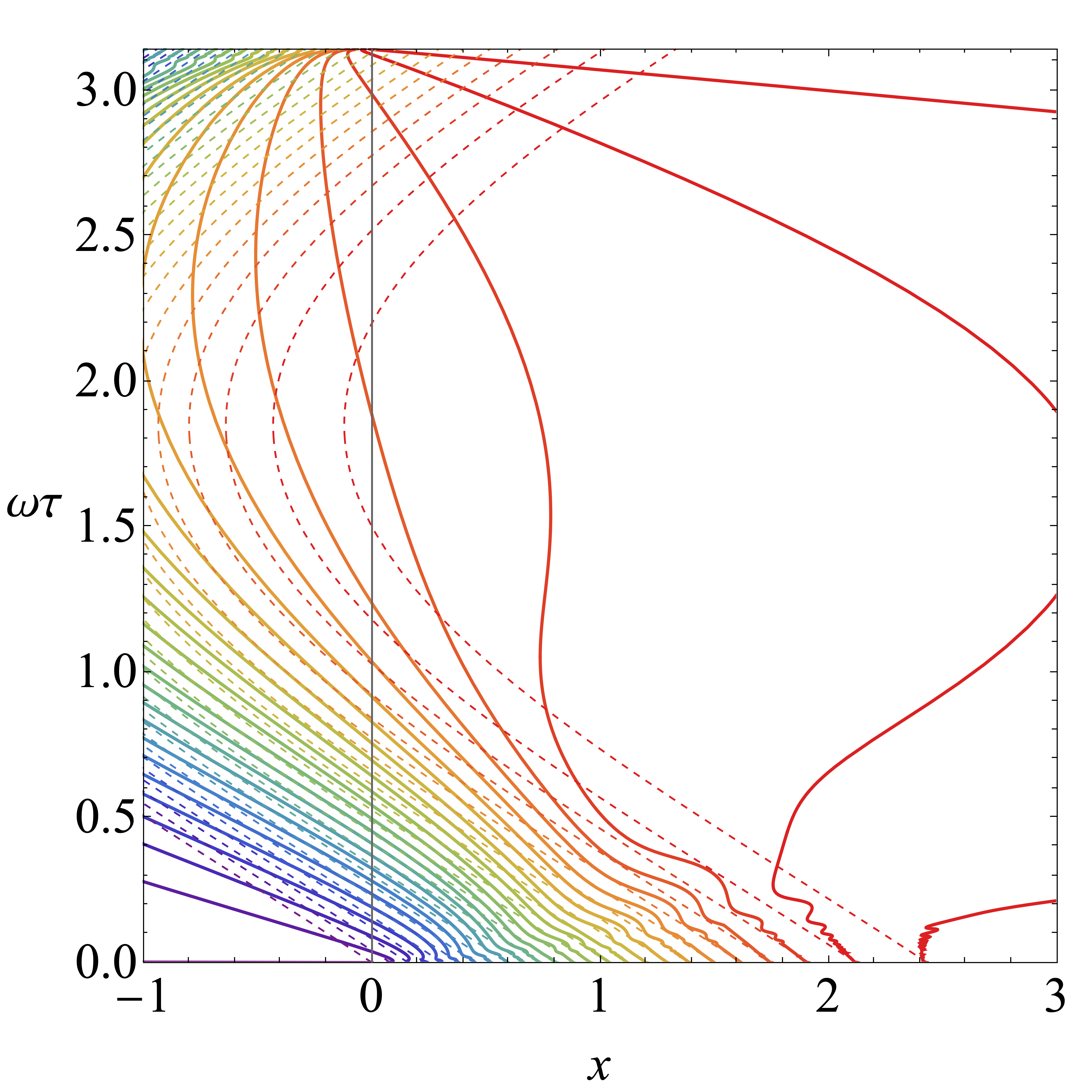}}}%
	%\qquad
	\hspace{5mm}
	\subfloat[]{{\includegraphics[height=7.2cm]{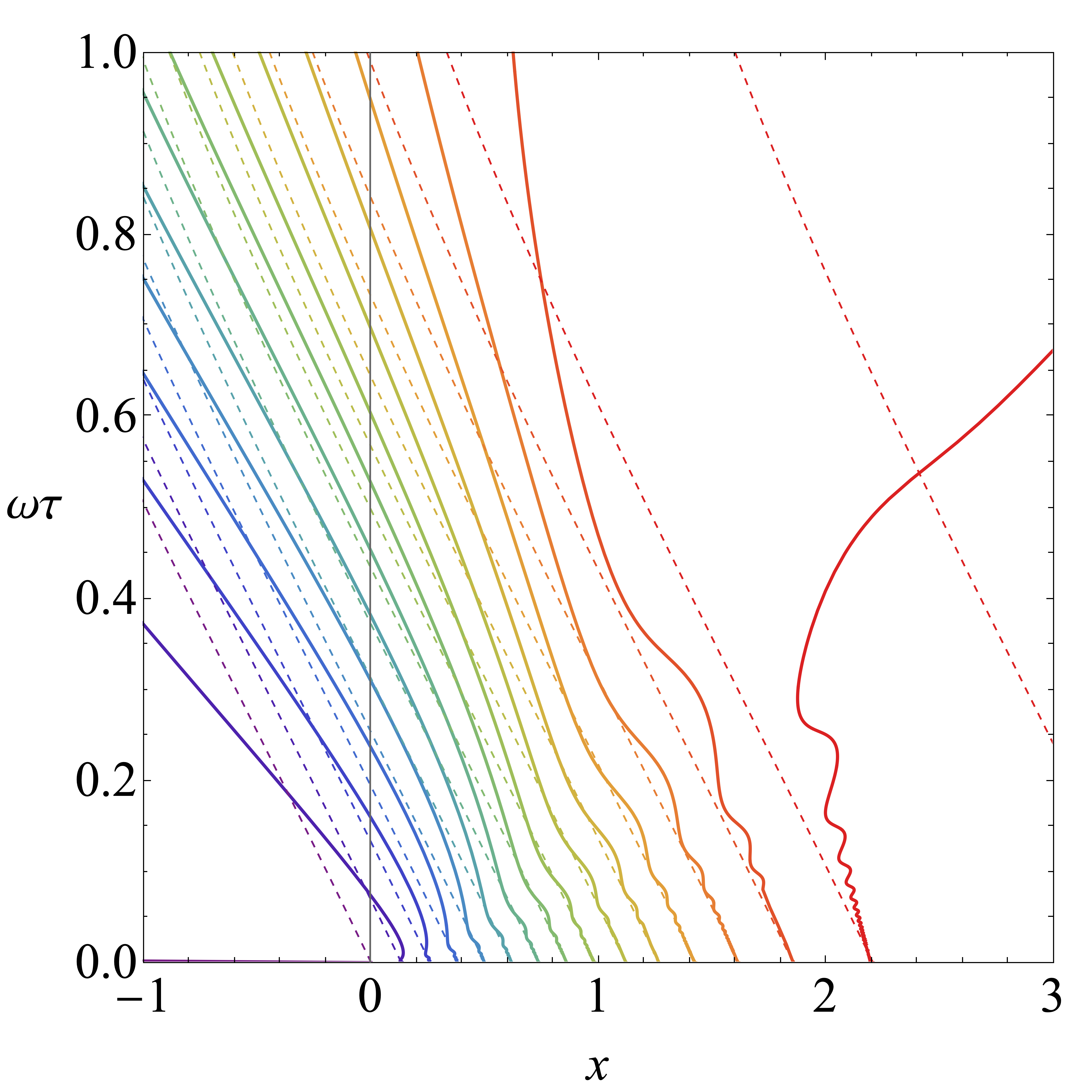}}}
	\caption{The Bohm trajectories for case of the particle being initially found on the right hand side of the axis.  In (a), trajectories are separated such that two adjacent lines bound the evolution of $6.67\%$ of the probability density, and in (b) they bound $10\%$ of the probability density.}%
	\label{fig:bts}%
\end{figure}

In Fig.~\ref{fig:bts}, we show the Bohm trajectories for the state with $x_0=0.55$, $p_0=-1.925$, initially found on the right hand side of the well.  This corresponds to the behaviour of the current $J_+(x,t)$ from the previous section.  Looking at the zoom of the trajectories in Fig.~\ref{fig:bts}(b), we can observe the same behaviour that is seen in the Moshinsky case -- initially an anti-Zeno effect, during which the trajectories exit faster than they would classically, followed by a Zeno effect a short while later, where trajectories exit more slowly than in the classical case.

This lines up with the behaviour of $\mathbb{J}_+(t)-J_+(t)$ displayed in Fig.~\ref{fig:currs}(b), and is hence a representation on the trajectory level of the source of the LG2 violations.

\subsection{Wigner Function Approach}
\begin{figure}
	\includegraphics[height=9.4cm]{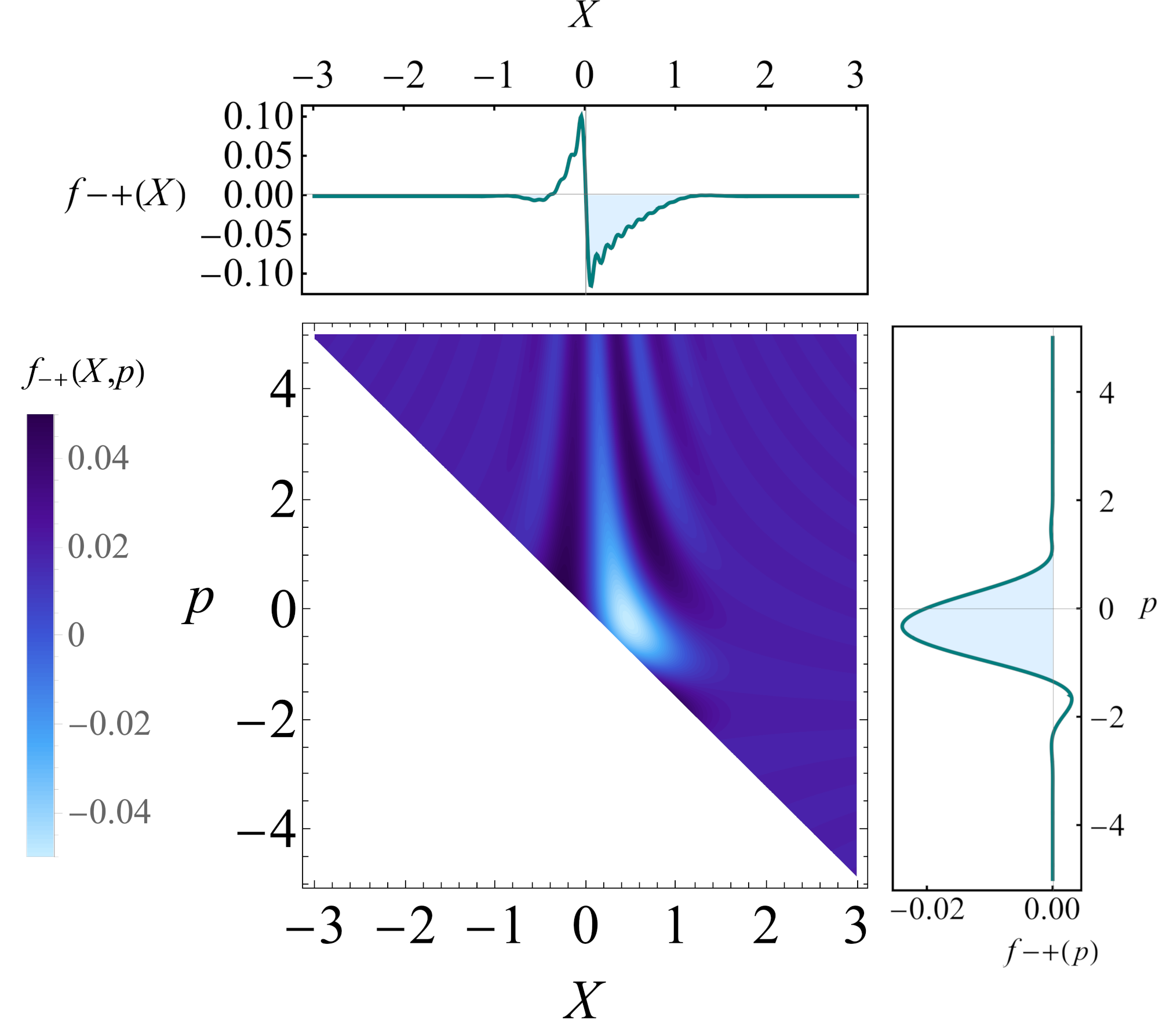}
	\caption{We plot the phase-space density $f_{-,+}(X,p)$, in the case corresponding to an LG2 violation of $-0.113$. Inset are the $X$ and $p$ marginals, shaded over regions of negativity.}%
	\label{fig:wig}%
\end{figure}

Another way to understand the LG2 violations, is within the Wigner representation \cite{wigner1932a, hillery1984a, tatarskii1983a, case2008, halliwell1993a}.  Since coherent states have non-negative Wigner functions, the source of MR violation lies in the non-gaussianity of the Wigner transform of the operators describing the measurement procedure.  We calculate these transformations in Appendix ~\ref{app:wig}, which 
allows us to write the quasi-probability as a phase-space integral,
\begin{equation}
	q(s_1,s_2)=\int_{-\infty}^{\infty}\mathop{dX}\int_{-\infty}^{\infty}\mathop{dp}f_{s_1, s_2}(X,p)
\end{equation}
with the phase-space density $f_{s_1, s_2}(X,p)$ given by
\begin{equation}
\label{eq:wigfin}
f_{s_1, s_2}(X,p)=\frac12 W_{\rho}(X,p)\left(1+\Re\erf\left(i(p-p_0)+s_1 X\right)\right)\theta(s_2 X_{-\tau}),
\end{equation}
where $W_\rho(X,p)$ is the Wigner function of the initial state. In Fig.~\ref{fig:wig}, we plot this phase-space density $f_{-+}(X,p)$ for the state $x_0=0.55$, $p_0=-1.925$ and $\omega\tau=0.55$.  To make it clear that this integrates to a negative number, we numerically determine the marginals $f_{-+}(p) = \int_{-\infty}^{\infty}\mathop{dX}f_{-,+}(X,p)$ and likewise $f_{-+}(X)=\int_{-\infty}^{\infty}\mathop{dp}f_{-,+}(X,p)$, plotting them as insets in Fig.~\ref{fig:wig}.  It is clear from a simple inspection of these marginals that they will integrate to a negative number.

In Appendix~\ref{app:wig}, we plot the intermediate result Eq.~(\ref{eq:wigint}), where it is apparent how the choice of second measurement hones in on the negativity introduced by the initial measurement, ultimately leading to the LG2 violations in Section~\ref{sec:LGresults}.

\section{Modified Frameworks}
\label{sec:mf}
In this section, we broadly generalise our analysis, finding larger violations using the Wigner variant of the LG2s, and by analysing different measurements, and extending the analysis to squeezed and thermal coherent states.
\subsection{Achieving larger violations using the Wigner LG2 inequalities}
\label{subsec:wiglg}
Given the modest size of the LG violations obtained for a single gaussian, compared to the L{\"u}ders bound, it is natural to ask if there are modified situations in which larger violation can be obtained. One way of doing this is to examine slightly different types of inequalities known as the Wigner-Leggett-Garg (Wigner LG) inequalities \cite{saha2015, naikoo2020}. For the two-time case these arise as follows. The quasi-probability is readily rewritten as
\bea
q(s_1,s_2) &=& {\rm Re} \langle ( 1 - P_{-s_2} (t_2) ) ( 1 - P_{-s_1} (t_1) ) \rangle
\nonumber \\
&=& 1 - \langle P_{-s_1} (t_1)  \rangle -  \langle P_{-s_2} (t_2)  \rangle  + q(-s_1, -s_2).
\eea
However, from a macrorealistic perspective, there is nothing against considering the similar quasi-probability
\beq
\label{eqn:qw}
q^W (s_1,s_2) = 1 - \langle P_{-s_1} (t_1)  \rangle -  \langle P_{-s_2} (t_2)  \rangle 
+ p_{12} (-s_1,-s_2),
\eeq
(where recall $p_{12}$ is the sequential measurement probability Eq.~(\ref{eq:seqprob}))
since the two are the same classically. The relation $q^W (s_1,s_2) \ge 0 $ is  a set of Wigner LG2 inequalities (recalling the factor of $\tfrac{1}{4}$ difference between an LG2 and a QP). It differs from the usual LG2 inequalities by the presence of interference terms, which can be positive or negative, which indicates that violations larger the usual L\"uders bound (on the QP) of $- \frac{1}{8}$ might be obtained. The difference between them from an experimental point of view is that the original quasi-probability is measured from three different experiments (determining $\langle Q_1 \rangle$, $ \langle Q_2 \rangle $ and $C_{12}$) but the sequential measurement formula appearing in the Wigner version is measured in a single experiment.

To get a sense of how much larger the maximum violation might be, we take the simple case of one-dimensional projectors $P_{-s_1} (t_1) = | A \rangle \langle A |$ and 
$P_{-s_2}(t_2) = | B \rangle \langle B | $ and we find
 \beq
q^W (A,B) = 1 - | \langle \psi | A \rangle |^2  - | \langle \psi | B \rangle |^2 + 
| \langle \psi | A \rangle |^2  | \langle A | B \rangle |^2.
\label{qWAB}
\eeq
Simple algebra reveals the lower bound as  $ - \frac{1}{3}$, which is achieved with $ \langle A | B \rangle = 1 / \sqrt{3} $ and $ | \psi \rangle = (1 / \sqrt{6} ) ( | A \rangle + \sqrt{3} |B \rangle $). (It seems like that this is the most negative lower bound for all possible choices of projection but we have not proved this.)
This bound is significantly larger than the 
usual L\"uders bound on the QP of $ - \tfrac{1}{8}$.

In Section 3 we found that the quasi-probability $q(-,+)$ gives the greatest negativity so we compare with the corresponding Wigner expression,
\bea
q^W (-,+) &=&1 - \langle P_{+} (t_1)  \rangle -  \langle P_{-} (t_2)  \rangle 
+ p_{12} (+,-)
\nonumber \\
&=& q(-,+) + \left[ p_{12}(+,-) - q(+,-) \right],
\eea
where we have made use of Eq.~(\ref{eqn:qw}) for $q(-,+)$
We first note that Eq.~(\ref{eq:qpdiffs}) may be written $q(-,+) = p_{12} (-,+) + I$, where $I$ denotes the interference terms (i.e. the difference between the classical and quantum currents, the second term in Eq.~\ref{eq:qpdiffs}).
The analogous relations for $q(+,-)$ is readily derived and we find $ q(+,-) = p_{12}(+,-) - I $. (The difference in sign is expected on general grounds \cite{halliwell2016b}). We thus find
\beq
q^W (-,+) = p_{12} (-,+) + 2 I,
\eeq
so the interference term producing the violations is twice as large as the one in $q(-,+)$.

\begin{figure}
	\includegraphics[height=5.4cm]{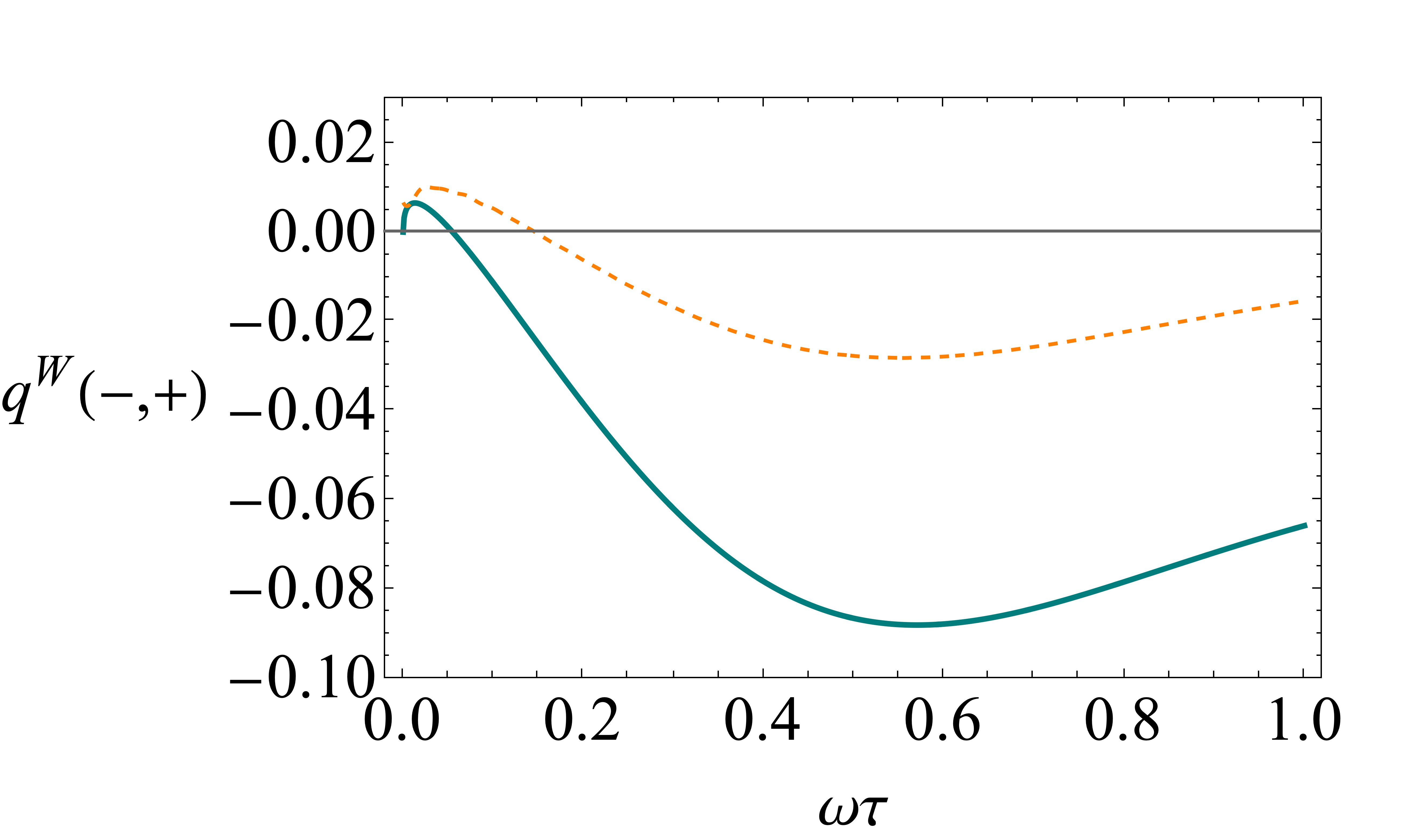}
	\caption{Plot of $q^W(-,+)$ for the coherent state $x_0=-0.55$, $p_0=-1.925$,  as well as the standard quasi-probability (dashed), i.e. the LG2 in Fig.~\ref{fig:lg2}(b) times a factor of $\frac14$.}%
	\label{fig:wigqp}%
\end{figure}

The computation of $p_{12} (-,+)$ can be carried out by integrating the chopped current $J_-(t)$, Eq.~(\ref{eq:chopcurr}) in Section 4.  Using the maximally violating state found in Section~\ref{sec:LGresults}, we find a largest violation of $-0.0881$, approximately three times larger than the standard LG2 violation, as well as a larger fraction of the conjectured Wigner LG2 L{\"u}ders bound of $-\frac{1}{3}$.  This is plotted alongside the standard LG2 in Fig.~\ref{fig:wigqp}, where the violation is both larger in magnitude, and present for a larger range of measurement intervals.

%In fact, from the plot of the currents in Fig.~(\ref{fig:currs}) it is readily seen that this current is essentially zero for most of the range of interest which means that $p_{12} (-,+)$ will be approximately zero. It follows that we expect a Wigner LG2 violation approximately twice as large as the usual LG2 violation. 
%It will however be a  smaller fraction of the maximal Wigner LG2 violation, but it might be argued that what is most important here in the search for non-classical effects is the degree of departure for a genuine probability lying between $0$ and $1$ and the Wigner expression gives the larger departure.

As an aside, we note an interesting aspect of Eq.~(\ref{qWAB}), which is that the last term, corresponding to the sequential measurement probability, factors in two parts. This factoring will also hold for for more general projections at $t_2$ as long as the projection at $t_1$ is one-dimensional. This may have some advantages in terms of meeting the non-invasiveness requirement on the measurements. 
It seems plausible that one could find macrorealistic arguments implying that the sequential measurement probability factors. Then the first factor is the probability of finding $ |A \rangle$ in an initial state $| \psi \rangle $ and the second factor is the probability of finding $|B \rangle$ when the systems is prepared in state $|A\rangle$. These quantities could therefore be obtained in two different experiments with two different preparations with just a single measurement in each, for which there is no issue with invasiveness.

The other obvious way of getting larger violations is to consider von Neumann measurements, which involves making finer-grained measurements than the simple dichotomic ones used here and then coarse graining the probability to compute the correlators \cite{pan2018,dakic2014, budroni2014,wang2017,kumari2018}.  For example, one could make measurements onto three regions of the $x$-axis, $ x<0$, $ 0 \le  x \le L$ and $ x> L$ at the first time and then coarse-grain the two-time probabilities into probabilities for the usual coarse graining $x<0$ and $x>0$.
This produces extra interference terms which can enhance the violations. For the LG3 inequalities, von Neumann measurements can produce violations up to the algebraic maximum of $-2$. For the LG2 inequalities, the enhancement is smaller since there is only one correlator and the LG2 violations can be no more than $-1$. This corresponds to $-\frac{1}{4}$ in the quasi-probability, which we see is not as big as the violation of $ - \frac{1}{3}$  that can be produced produced by the Wigner LG2.

\subsection{Coherent state projectors}
\label{subsec:cohproj}
It is useful to know what else may be possible beyond using $\theta(\hat x)$ projectors.  We note investigations \cite{bose2018} into smoothed $\theta(\hat x)$ measurements, showing LG violations persist under smoothing of measurements up to the characteristic length-scale of the oscillator \cite{mawby2022}.  Modular variables such as $\cos(k\hat x)$ have also been investigated, and readily produce significant LG violations \cite{asadian2014}. These examples show that the LG violations are not due to the sharpness of projective measurements with $\theta(\hat x)$.  

In this section we will look at tests of macrorealism using coherent state projectors \cite{yuen1978,yuen1983, dariano2003}, which are interesting since they leave the post-measurement state gaussian, and are easily experimentally realised.

We now consider coherent state projectors, where we have $P_+=\ketbra{\beta}$, and $P_-=\id -\ketbra{\beta}$ defining a dichotomic variable in the usual way through Eq.~(\ref{eq:ptoq}). The quasi-probability is given by \begin{equation}
q(+,+)=	\Re \ev{e^{iHt_2}\ketbra{\beta_2}e^{-iH\tau}\ketbra{\beta_1}e^{-iHt_1}}{\alpha},
\end{equation}
where we make two simplifying observations.  Firstly,  all the time evolution may be absorbed into the measurement projectors.  Secondly, without loss of generality, we work with $\alpha=0$, where the change in phase-space location may be absorbed into $\beta_1$ and $\beta_2$.  It is hence entirely equivalent to analyze
\begin{equation}
q(+,+)=\Re \braket{0}{\gamma_1}\!\!\braket{\gamma_1}{\gamma_2}\!\!\braket{\gamma_2}{0},
\end{equation}
with the relation $\gamma_i=e^{-i\omega t_i}\beta_i-\alpha$.
The overlap between two coherent states is given by
\begin{equation}
\braket{\beta}{\alpha}=e^{-\frac12(\abs{\alpha}^2+\abs{\beta}^2-2\alpha \beta^*)}, 
\end{equation}
and we readily find
\begin{align}
q(+,+)&=\exp(-\lvert\gamma_1\rvert^2 -\lvert \gamma_2\rvert^2)\Re \exp(\gamma_1 \gamma_2^*),\\
q(+,-)&=\exp(-\lvert\gamma_1^2\rvert)\left(1-\Re\exp(\gamma_1 \gamma_2^* -\lvert\gamma_2\rvert^2)\right),
\end{align}
where $q(-,+)$ is found by a relabelling, and $q(-,-)$ does not lead to any violations.
To determine the largest violations, it is useful to note these quasi-probabilities depend only on the magnitude of $\gamma_1$ and $\gamma_2$, and the phase difference between them.  

In $q(+,+)$, $\lvert \gamma_1\rvert$ and $\lvert \gamma_2 \rvert$ appear in the same way, so we set them to be equal, and find a largest violation of $-0.0133$ at $\gamma_1=1.55$, $\gamma_2=1.55 e^{-1.047 i}$, which is about $10\%$ of the maximal violation.

For $q(-,+)$, since the violation is aided by the negative sign on $\Re \exp(\gamma_1 \gamma_2^* -\lvert\gamma_2\rvert^2)$, it is easy to see the largest violation will occur when both $\gamma_1$ and $\gamma_2$ are purely real.  We readily find that the largest violation is approximately $-0.1054$ with $\gamma_2=\frac12\gamma_1=0.536$, which is about $84\%$ of the maximum.

Violations meeting the L{\"u}ders bound may be achieved if a superposition state is chosen which satisfies Eq.~(\ref{eq:maxQP}.  The superposition state $\ket\psi = -\ket{\beta_1} -\ket{\beta_2}$ is properly normalized and gives a maximal violation for $q(+,+)$ if the coherent states are chosen so that $\braket{\beta_1}{\beta_2} = -\frac12$.  Similarly for $q(+,-)$, the state $\ket\psi = \ket{\beta_1} - \sqrt{3}\ket{\beta_2}$ leads to a maximal violation if we choose $\braket{\beta_1}{\beta_2}=\frac{\sqrt{3}}{2}$.
\subsection{Squeezed States}
The squeezed coherent state may be written~\cite{schleich2001},
\begin{equation}
\ket{\alpha,\zeta}=D(\alpha)S(\zeta)\ket0.
\end{equation}
While the squeezing operator $S(\zeta)$ does not commute with the displacement operator $D(\alpha)$, there is a simple braiding relation, allowing us to write
\begin{equation}
	\ket{\alpha,\zeta}=S(\zeta)D(\beta)\ket0=S(\zeta)\ket\beta,
\end{equation}
with $\beta$ depending on both $\alpha$ and $\zeta$.  With the quasi-probability given by
\begin{equation}
q(+,+)=\Re \ev{\theta(\hat x)\theta(\hat x(t))}{\psi},	
\end{equation}
for $\ket\psi$ given by a squeezed coherent state. We can consider moving the $S(\zeta)$ in $\ket\psi$ onto each $\theta(\hat x)$ function, resulting in $S^\dagger(\zeta)\theta(\hat x)S(\zeta)$ twice.  Since the squeezing operator has the action of a canonical transform, taking $\hat x$ and $\hat p$ into a linear combination of themselves, we have that 
\begin{equation}
S^\dagger(\zeta)\theta(\hat x)\theta(\hat x (t))S(\zeta)=\theta(a\hat x + b\hat p)\theta(c\hat x + d\hat ),
\end{equation}
for some $a,b,c,d$ that may depend on $t$. 
We now note that $a\hat x + b\hat p$ may be written as $\lambda(\hat x \cos(t')+\hat p \sin(t'))$ for some $\lambda> 0$ and some $t'$, and since the theta-function is invariant under scaling, we see that
\begin{equation}
	S^\dagger(\zeta)\theta(\hat x)\theta(\hat x (t))S(\zeta)=\theta(\hat x(t_1'))\theta(\hat x(t_2')).
\end{equation}
This means the QP for a squeezed coherent state is equal to the QP for some other coherent state $\beta$, with different measurement times $t_1'$, $t_2'$.  Hence the operation of squeezing will not increase the largest possible violation reported in Section~\ref{sec:LGresults}, although for certain states with sub-optimal violation, squeezing can increase the amount of violation. 
\subsection{Thermal States}
The thermal coherent state at a temperature $T$ is given by
\begin{equation}
	\rho_{\text{th}}(\alpha, T)=\frac{1}{Z}\sum_{n=0}^{\infty}e^{-\frac{n\hbar\omega}{k_B T}} \ketbra{n,\alpha},
\end{equation}
where $k_B$ is the Boltzmann constant, $\ket{n,\alpha}$ are energy eigenstates displaced by $\alpha$ in phase-space~\cite{oz-vogt1991}. The partition function $Z$ is given by
\begin{equation}
Z=\frac{1}{1-e^{-\frac{\hbar \omega}{k_B T}}}.
\end{equation}
Since this state is a mixture, it is simple to update the calculation Eq.~(\ref{eqn:quassum}) to using this state, leading to

\begin{equation}
q(+,-)=-\frac{1}{Z}\Re\sum_{\ell=0}^{\infty}e^{-\frac{\ell\hbar\omega}{k_B T}} \sum_{n=0}^\infty e^{-i(n-\ell)\omega\tau}J_{\ell n}(x_1,\infty)J_{\ell n}(x_2,\infty)
\end{equation}
where the $J_{n\ell}$ matrices are given by Eq.~(\ref{eq:Jmn}), except for the cases $n=\ell$, where they must be calculated explicitly.  A similar result may be calculated for the correlators using Eq.~(\ref{eqn:corr}), allowing the analysis of LG3 and LG4 inequalities.

\begin{figure}
\includegraphics[height=5.3cm]{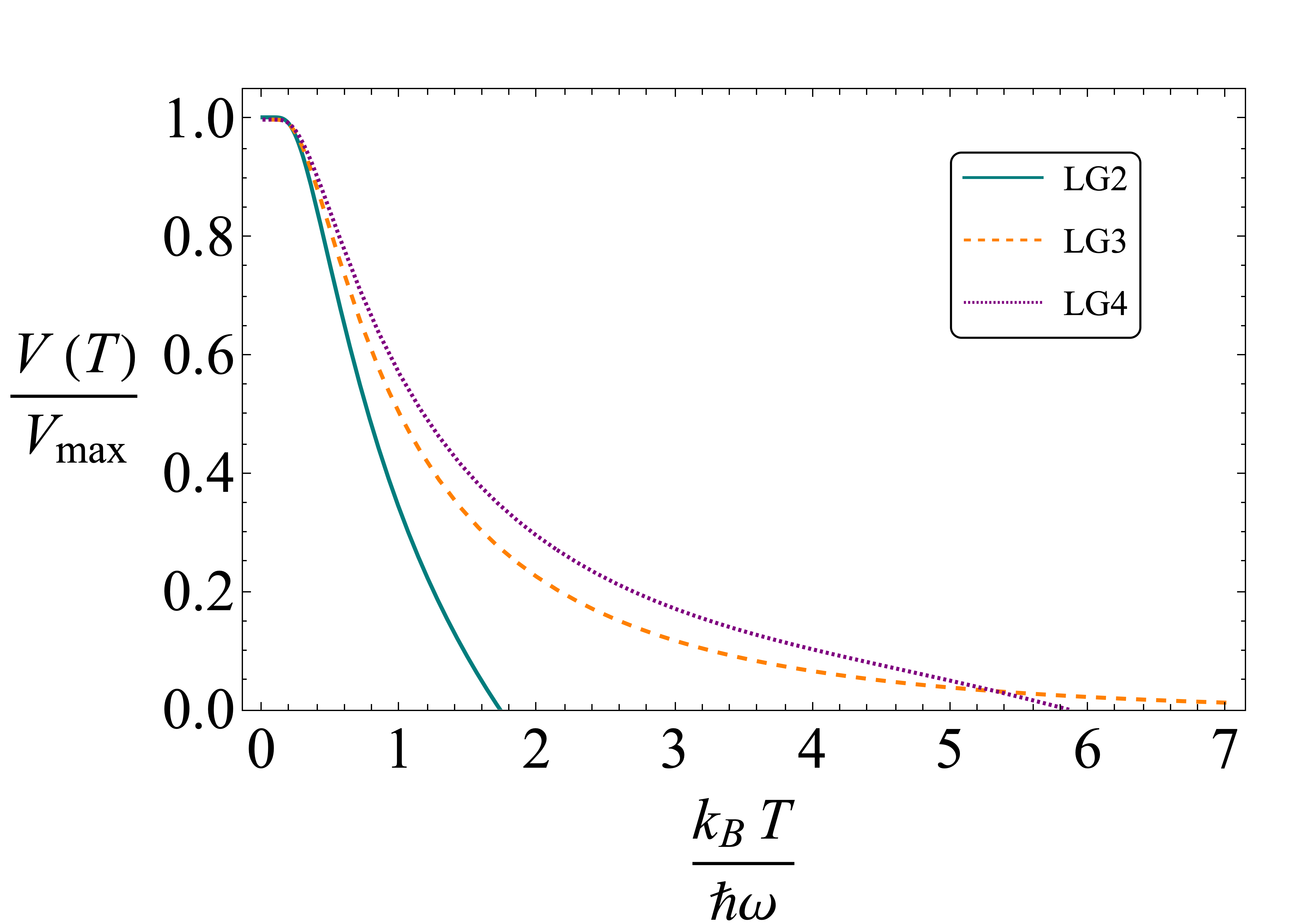}
	\caption{The LG2, LG3 and LG4 violations are plotted as a fraction of their largest violation, for a state and time which realises $V_{\text{max}}$, with varying temperature.}%%
	\label{fig:thermLG}
\end{figure}

In Fig.~\ref{fig:thermLG} using the states found in Section~\ref{sec:LGresults}, we plot the behaviour of the largest violation, as temperature is increased.  We see the violation persists up to temperatures $k_B T\approx \hbar \omega$, with some preliminary evidence that LG3 and LG4 violations may be more robust against thermal fluctuations in the initial state.
\section{Summary}
\label{sec:sum}
We have undertaken a study of LG violations in the quantum harmonic oscillator for a dichotomic variable $Q = \text{sgn}(x)$ and for an initial state given by a coherent state and closely related states. In Section \ref{sec:corecalc}, building on our earlier work with energy eigenstates of the QHO \cite{mawby2022}, we showed how the quasi-probability, and hence the temporal correlators, may be expressed as a discrete infinite sum which is amenable to numerical analysis. We applied this analysis to the LG2, LG3 and LG4 inequalities and carried out parameter space searches. We found LG violations of magnitude 22\%, 28\% and 26\% of the maximum possible for the LG2, LG3 and LG4 inequalities, respectively, and gave the specific parameters for which these violations are achieved.  These violations appear to be robust under small parameter adjustments.

The LG2 violation in the case $x_0=0$ agrees with that reported in Ref.~\cite{halliwell2021}. The LG4 violation is significantly smaller than that reported in Ref.~\cite{bose2018}, which used a coherent state with large momentum, and in fact we found no violations in that regime, although the authors note it involves a very narrow parameter range.

In Section \ref{sec:physmec} we sought a physical understanding of the mechanism producing the violations. We showed how to relate the quasi-probability (LG2) to a set of currents for projected initial states. We calculated and plotted these currents and also plotted their associated Bohm trajectories along with their classical counterparts. The plots showed the clear departures from classicality and give both a visual understanding and independent check of the LG2 violations described in Section \ref{sec:corecalc}.  We also provide a small-time expansion for the LG2s, which is valid for general states. We noted that the quantum effect producing the violations is essentially the diffraction in time effect first noted by Moshinsky~\cite{moshinsky1952,moshinsky1976}.

We explored the same issues from a different angle using the Wigner representation. The Wigner function of the initial coherent state is everywhere non-negative. We determined and plotted the Wigner function of the chopped initial state appearing in the quasi-probability. It has significant regions of negativity which are clearly the source of the LG2 violation.

In Section \ref{subsec:wiglg}, we extended our results to the slightly different Wigner LG inequalities, which are phrased in terms of the sequential measurement probability, and allow for larger violations, where we found a two-time violation three times greater in magnitude than the standard LG2s.  We also noted it is likely possible to increase the LG violations through the use of von Neumann measurements.   In addition we noted a possible advantage of the Wigner LG2 inequalities, in some cases, in terms of meeting the non-invasiveness requirement.	

We briefly noted in \ref{subsec:cohproj} that our work is readily generalized from pure projective measurements to smoothed step function projectors, gaussian projectors and modular variables such as $\cos ( \hat x) $. We also examined the LG2 inequality for the case in which both projectors are taken to be projections onto coherent states. We showed that decent violations are possible for an initial coherent state and that a maximal violation arises when the initial state is a superposition of two coherent states.

Finally we finished Section \ref{sec:mf} briefly discussing how the LG violations may be modified using families of states similar to a coherent state. We showed that the QP for any squeezed state is equal to the QP for some other coherent state, hence squeezing will not increase the largest violation found, however for a state with sub-optimal violation it may improve the violation.  We also considered a thermal initial state and estimated the degree to which thermal fluctuations may affect the degree of violation.
\section*{Acknowledgements}	
We are grateful to Sougato Bose, Dipankar Home, and Debarshi Das for useful discussions.  We also thank Michael Vanner for discussions on the experimental feasibility of the tests discussed in this paper.
\appendix
\section{Calculation of correlators}
\label{app:timeev}
The two-time quasi-probability for a coherent state with a generic position basis measurement $m(\hat x)$ is defined,
\begin{equation}
q(+,+)=\Re\ev{e^{\frac{i H t_2}{\hbar}}m(\hat x)e^{-\frac{i H \tau}{\hbar}}m(\hat x)e^{-\frac{i H t_1}{\hbar}}}{\alpha}.
\end{equation}
We are primarily interested in the case $m(\hat x) = \theta(\hat x)$, but what follows holds for more general $m(\hat x)$, e.g gaussian measurements.  Writing this in terms of the displacement operator, we have
\begin{equation}
q(+,+)=\Re\ev{D^{\dagger}(\alpha)e^{\frac{i H t_2}{\hbar}}m(\hat x)e^{-\frac{i H \tau}{\hbar}}m(\hat x)e^{-\frac{i H t_1}{\hbar}}D(\alpha)}{0}.
\end{equation}

Since the displacement operator is unitary, we have $D^\dagger(\alpha) D(\alpha)=\mathds{1}$. Hence if we may commute the two displacement operators to be neighbours, we will clearly reach a vast simplification of the calculation.

To make the exposition clearer, we consider splitting this expression into two states
\begin{align}
\ket{M(\alpha, t_1, \tau)}&=e^{-i\hat H \tau}m(\hat x)e^{-\frac{iHt_1}{\hbar}}D(\alpha) \ket0,\\
\bra{M(\alpha, t_2, 0)}&=\bra0 D^\dagger(\alpha)e^{\frac{i H t_2}{\hbar}}m(\hat x),
\end{align}
where we then have $q(+,+)=\Re \braket{M(\alpha, t_2, 0)}{M(\alpha, t_1, \tau)}$, where we have introduced the notation $\ket{M(\alpha, t, s)}$ to represent the coherent state measured with $m(\hat x)$ at time $t$, then evolved by time $s$.

Considering now the displacement operator acting to the left, we write $m(\hat x)D(\alpha)=D(\alpha)D^\dagger(\alpha)m(\hat x)D(\alpha)=D(\alpha)m(\hat x + x_\alpha)$, with $x_\alpha=\sqrt2 \Re \alpha$. We then have 

\begin{equation}
\ket{M(\alpha, t_1, \tau)} = e^{-\frac{i\omega t_1}{2}}e^{-i \hat H \tau}D(\alpha(t_1))m(\hat x + x_1)\ket 0.
\end{equation}
Using the standard result that $e^{-i H t}D(\alpha)e^{i H t}=D(\alpha(t))$, we can rewrite this as
\begin{equation}
\ket{M(\alpha, t_1, \tau)} =e^{-\frac{i\omega t_1}{2}}D(\alpha(t_2))e^{-i H\tau}m(\hat x + x_1)\ket0.
\end{equation}
This says that the post-measurement state is the evolution of the regular ground-state undergone a translated measurement, translated by the displacement operator, to a classical trajectory.  Proceeding similarly with the other term, we find
\begin{equation}
\bra{M(\alpha, t_2, 0)}=e^{\frac{i\omega t_2}{2}}\bra{0} m(\hat x + x_2)D^\dagger(\alpha(t_2)).
\end{equation}
Finally, contracting the two terms we are able to exploit the unitarity of $D(\alpha)$ to find
\begin{equation}
q(+,+)=\Re e^{\frac{i\omega \tau}{2}} \ev{m(\hat x + x_2)e^{-iH\tau}m(\hat x + x_1)}{0},
\end{equation}
\noindent
A calculation similar to that in our earlier paper Ref.~\cite{mawby2022} shows the quasi-probability is
\begin{equation}
q(+,+)=\Re e^{\frac{i\omega \tau}{2}}\sum_{n=0}^{\infty}e^{-i\omega \tau(n+\frac12)}\mel{0}{\theta(\hat x + x_2)}{n}\!\!\mel{n}{\theta(\hat x + x_1)}{0},
\end{equation}
and similarly for the other three components.  The matrix elements here are given by the $J_{mn}$ matrices from our earlier paper \cite{mawby2022,moriconi2007a},
\begin{equation}
J_{mn}(x_1,x_2)=\int_{x_1}^{x_2}\mathop{dx}\braket{m}{x}\braket{x}{n}.
\end{equation}
The quasi-probability is then
\begin{equation}
\label{eqn:quassum}
q(+,+)=\Re \sum_{n=0}^\infty e^{-in\omega\tau}J_{0n}(x_1,\infty)J_{0n}(x_2,\infty)
\end{equation}
For $m\neq n$, the $J$ matrices take the value
\begin{equation}
\label{eq:Jmn}
J_{mn}(x_1, x_2)=\frac{1}{2(\varepsilon_n- \varepsilon_m)}\left[\psi_m'(x_2)\psi_n(x_2)-\psi_n'(x_2)\psi_m(x_2)-\psi_m'(x_1)\psi_n(x_1)+\psi_n'(x_1)\psi_m(x_1)\right],
\end{equation}
where $\psi_n(x)=\braket{x}{n}$. For the $n=m=0$ case, the integration is completed manually yielding
\begin{equation}
J_{00}(x, \infty)=\frac12(1-\erf(x)).
\end{equation}
Hence writing out the quasi-probability with $n=0$ case of the sum handled, we have
\begin{multline}
\label{eqn:cohquas}
q(s_1, s_2)= \frac{1}{4}\Bigg[1+ s_1 \erf(x_1) +s_2 \erf(x_2)+\\ s_1 s_2\Bigg( \erf(x_1)\erf(x_2)+ 4\sum_{n=1}^\infty\cos (n\omega \tau)J_{0n}(x_1,\infty)J_{0n}(x_2,\infty)\Bigg) \Bigg].
\end{multline}
Comparing to the moment expansion of the quasi-probability, we obtain the correlators
\begin{equation}
\label{eqn:corr}
C_{12}=\erf(x_1)\erf(x_2)+ 4\sum_{n=1}^\infty \cos (n\omega \tau) J_{0n}(x_1,\infty)J_{0n}(x_2,\infty).
\end{equation}
The infinite sum may be evaluated approximately using numerical methods, by summing up to a finite $n$.  This calculation matches the analytically calculated special case of $x_0=0$ given in Ref. \cite{halliwell2021}, and while it is possible to make an analytic calculation for the more general case, it turned out not to be as useful as the numerical evaluation.  The exact result is found in terms of Owen-T functions, but for complex arguments, which rendered the behaviour chaotic when computed \cite{owen1965}.

The only source of non-classicality here lies in the infinite sum, and  with the $J_{0n}$ matrices expressed in terms of the oscillator eigenstates, this means there is a double exponential suppression $e^{-x_1^2 -x_2^2}$ of this non-classical term. This corresponds to the requirement that at least two measurements must make a significant chop of the state, which fits the intuition that without significant chopping, there is no mystery attached to which side of the axis the particle may be found on.

\section{Determination of LG violations}
\label{app:LGresults}

\begin{figure}
	\subfloat[]{{\includegraphics[height=5.8cm]{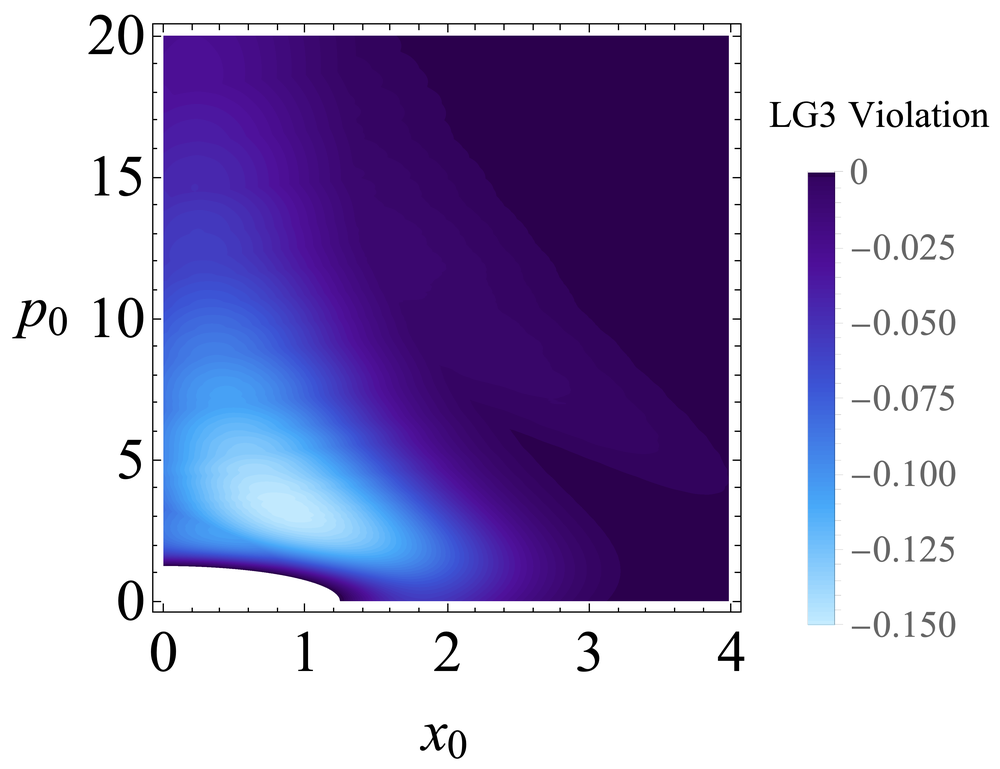}}}%
	%\qquad
	\hspace{5mm}
	\subfloat[]{{\includegraphics[height=5.8cm]{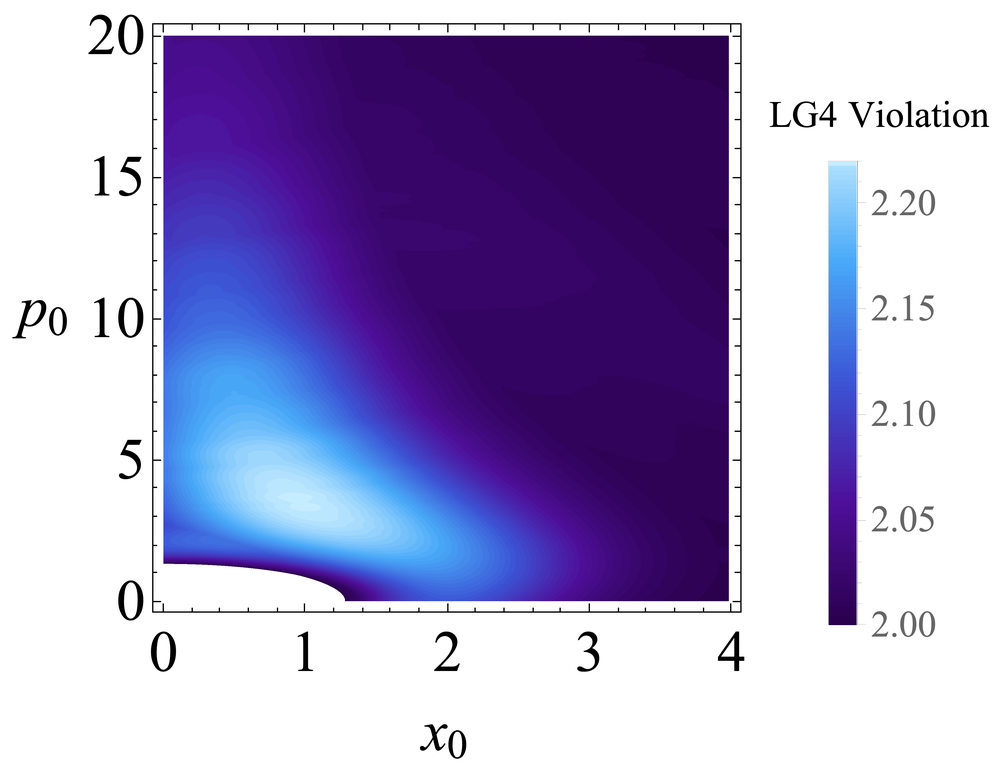}}}
	\caption{Plot (a) shows the greatest possible violations for the LG3 inequalities, plot (b) shows the same for the LG4 inequalities.}%
	\label{fig:param23}%
\end{figure}

In this appendix we fill in the details of the LG violations reported in Section \ref{sec:LGresults}.  Recall the variable parameters of the problem are $x_0$ and $p_0$, and the equal time spacing parameter $\tau$.

We also note, that it is sufficient to explore a single quadrant of the $x_0, p_0$ parameter space, which we take to be the positive quadrant.  For states with $x_0<0$, the quasi-probability may be recovered by inverting the sign of $s_1$.  Likewise for states with $p_0<0$, by allowing the interval between measurements to take values $0<\tau\leq 2\pi$, their behaviour is included in the positive quadrant.  This same argument applies to the LG inequalities in general, where their different permutations correspond to flips of measurement signs.

To represent the three-dimensional parameter space, for each $x_0$, $p_0$, we use numerical minimisation over $0<\tau\leq 2\pi$ to find the largest possible violation for that coherent state.  In this numerical procedure, we take the largest possible violation from all of the inequalities involved.

The results of this parameter space search for the LG3 and LG4 inequalities is shown in Fig.~\ref{fig:param23}, which shows similar behaviour to the LG2 inequality parameter space behaviour in Fig.~\ref{fig:lg2}.  As more measurement intervals are included in the LG tests, a broader range of states lead to violation.

LG tests on QHO coherent states are mathematically equivalent to LG tests on the pure ground state $\ket0$, which are we found to have violations only when at least one of the $\theta(\hat x)$ measurements involved is displaced from the axis by order of magnitude $1$.  Hence at the centre of each of these parameter space plots is the region where the coherent state is too similar to the ground state to have any LG violation.
\begin{figure}
	\subfloat[]{{\includegraphics[height=5.2cm]{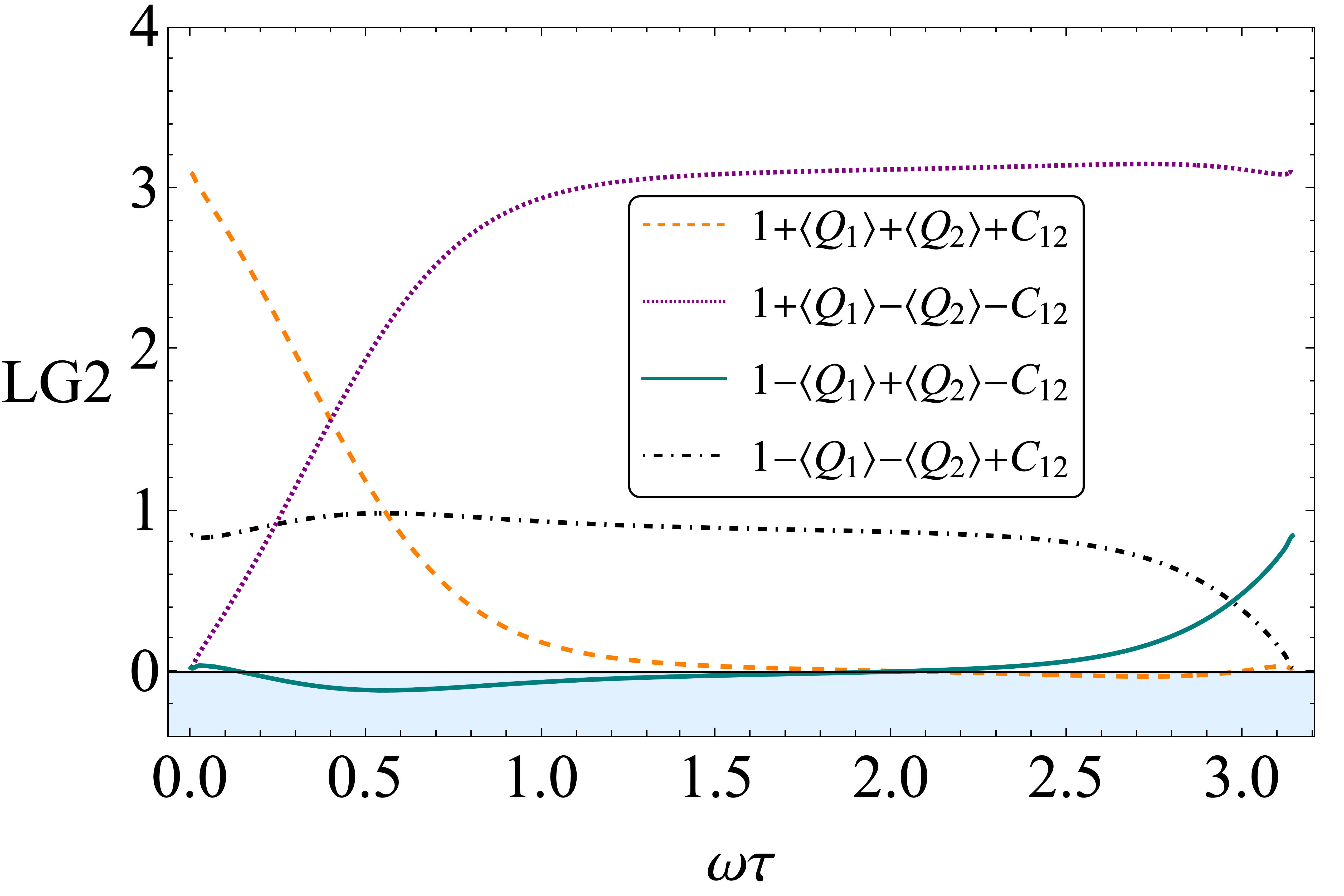}}}%
	%\qquad
	\hspace{5mm}
	\subfloat[]{{\includegraphics[height=5.2cm]{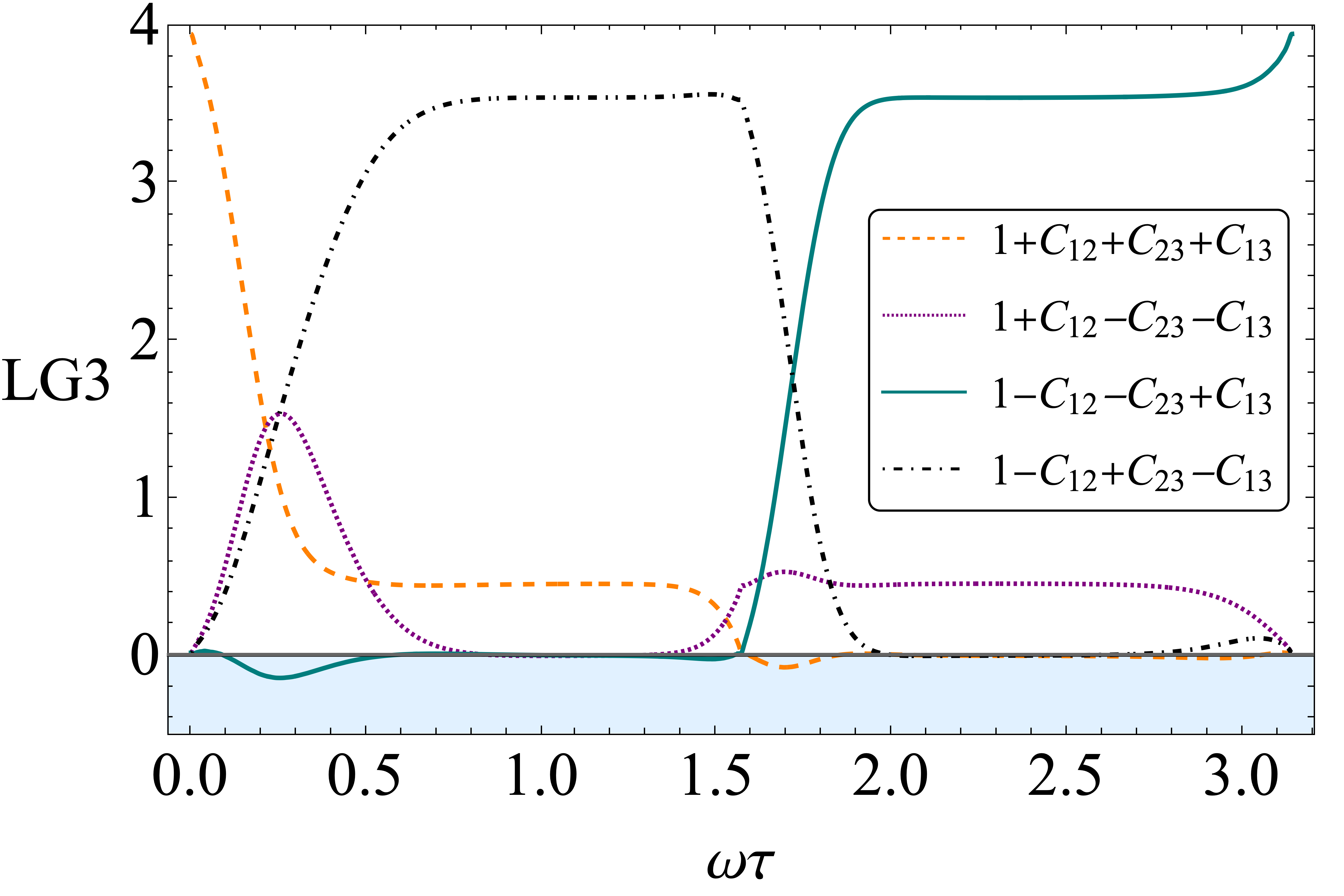}}}
	\caption{Plot (a) shows the four LG2 inequalities for the state with $x_0=0.55$, $p_0=-1.925$, which reaches a largest violation of $-0.113$ at  $\omega \tau = 0.555$.  Plot (b) shows the four LG3 inequalities for the state with $x_0=0.859$, $p_0=-3.317$, reaching a largest violation of $-0.141$ at $\omega\tau=0.254$. }%
	\label{fig:qpt}%
\end{figure}
\begin{figure}
	\subfloat[]{{\includegraphics[height=5.3cm]{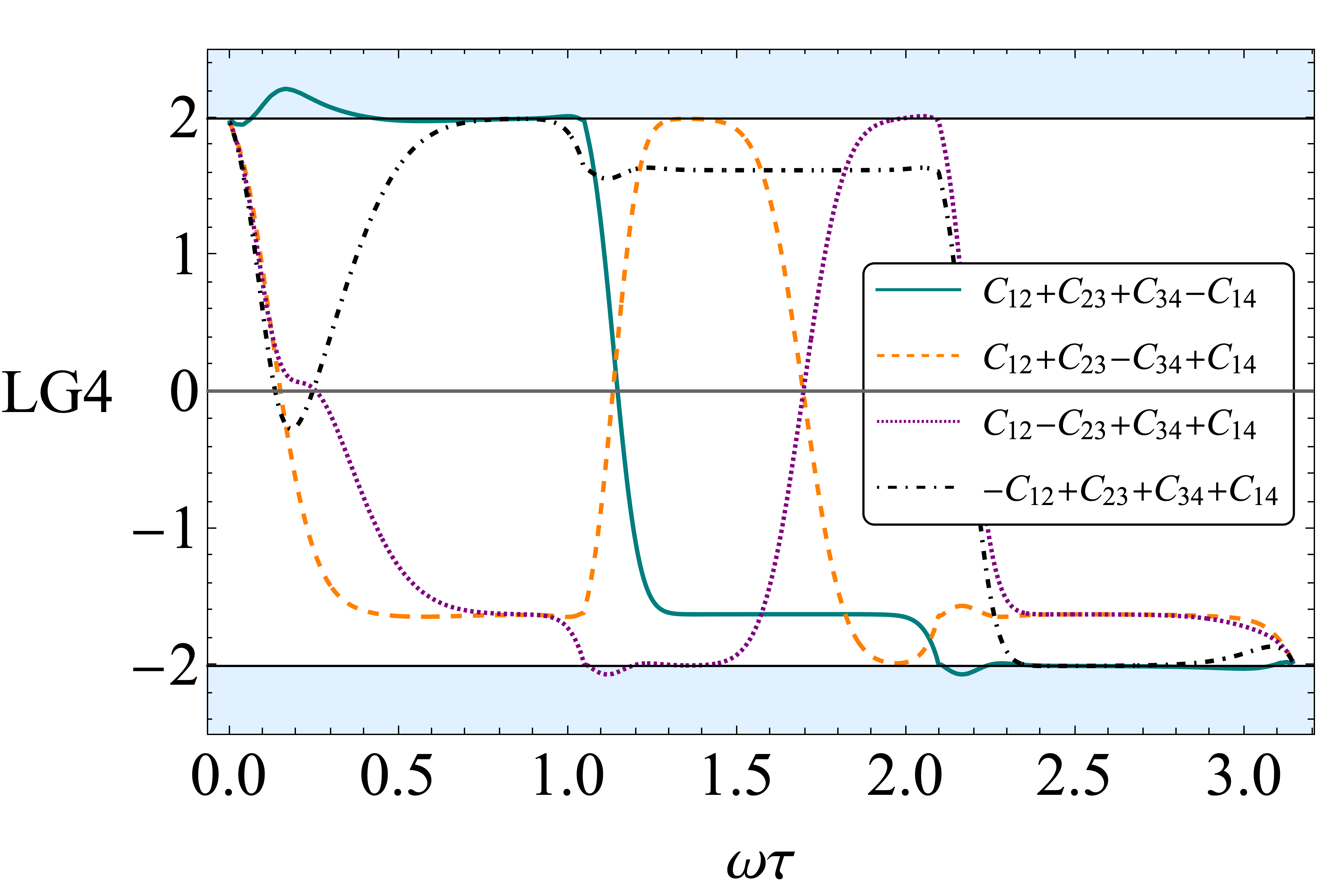}}}%
	%\qquad
	\caption{Plot of the four LG4 inequalities, for state $x_0=0.929$, $p_0=-3.666$, reaching a largest violation of $2.216$ at $\omega\tau=0.166$.}%
	\label{fig:LG4}%
\end{figure}

All the violations we have found are in states with initial position and momenta approximately on the length-scale of the width of the coherent state, $\sqrt{\hbar/(m\omega)}$.  However by appealing to Eq.~(\ref{eqn:corr}), we note that if one were to consider translating the measurement to $\theta(\hat x - x_i)$, the classical motion could be subtracted, and at least theoretically, the same magnitude of violations would exist for arbitrarily high $x_0$ and $p_0$.

In Figures \ref{fig:qpt} and \ref{fig:LG4}, we plot the temporal behaviour of the LG2s, LG3s and LG4s respectively, for the case in which the parameters are chosen to give the largest violation.

\section{Currents Analysis}
\subsection{Classical analogues}
\label{app:classan}
To understand the connection between the negativity of the quasi-probability Eq.~(\ref{eq:qpJ}) and the behaviour of the currents, it is very convenient to consider the analogous classical currents which are in general defined by
\begin{equation}
	\mathbb{J}(t)=\int_{-\infty}^{\infty}\mathop{dp}\int_{-\infty}^{\infty}\mathop{dx} p(t) \delta(x(t))w(x,p),
\end{equation}
for a suitably chosen initial phase-space distribution $w(x,p)$.  For the un-chopped current $\mathbb{J}(t)$ this is taken to be the Wigner function of the coherent state, $W(x,p,x_0, p_0)$ Eq.(\ref{eq:classWig}), which conveniently, is non-negative.  For the chopped curents it is taken to be $\theta(\pm x) W(x,p,x_0,p_0)$.  We then easily see that 
\begin{equation}
\label{eq:class}
	J(t)=\mathbb{J}_-(t)+\mathbb{J}_{+}(t),
\end{equation}
where $\mathbb{J}_{\pm}(t)$ are the classical analogues to the post-measurement currents, where since we have used the coherent state Wigner function, we have $\mathbb{J}(t)=J(t)$.

We begin by writing the classical phase-space density for the Gaussian state 
\begin{equation}
\label{eq:classWig}
\mathbb{W}(X,p, x_0, p_0)=\frac{1}{\pi}\exp(-(X -x_0)^2 -(p-p_0)^2),
\end{equation}
where harmonic time-evolution leads to rigid rotation in phase-space,
\begin{equation}
	\mathbb{W}(X,p,x_0,p_0,t)=\mathbb{W}(X \cos\omega t -p \sin\omega t,p \cos \omega t +X \sin \omega t, x_0, p_0).
\end{equation}
We have a similar result for the measured classical state, with
\begin{equation}
\mathbb{W}_{\pm}(X,p, x_0, p_0)=\frac{1}{\pi}\theta(\pm X)\exp(-(X -x_0)^2 -(p-p_0)^2),
\end{equation}
and
\begin{equation}
	\mathbb{W}_{\pm}(X,p,x_0,p_0,t)=\mathbb{W}_{\pm}(X \cos\omega t -p \sin\omega t,p \cos \omega t +X \sin \omega t, x_0, p_0).
\end{equation}
The chopped classical current is given by
\begin{equation}
\mathbb{J}_{\pm}	(x,t)=\int_{-\infty}^{\infty}\mathop{dp}\int_{-\infty}^{\infty}\mathop{dX}p\delta(X-x)\mathbb{W}_{\pm}(X,p,x_0, p_0,t)
\end{equation}
Completing the $X$ integral trivially, we have
\begin{equation}
\mathbb{J}_{\pm}(x,t)=\int_{-\infty}^{\infty}\mathop{dp} p\mathbb{W}_{\pm}(x,p,x_0, p_0,t).
\end{equation}
We are interested in the case of $x=0$, which we shorthand $\mathbb{J}_\pm(0,t)=\mathbb{J}_\pm(t)$, and is given by
\begin{equation}
	\mathbb{J}_\pm(t)=\frac{1}{\pi}\int_{-\infty}^{\infty}\mathop{dp} p\,\theta (\mp p \sin\omega t) \exp \left(-(p \cos\omega t-p_0)^2-(-p \sin \omega t-x_0)^2\right).
\end{equation}
The step-function here just flips the integral between the positive or negative half-plane, dependent on $\sgn(\mp \sin \omega t)=1$ and $\sgn(\mp \sin \omega t)=-1$ respectively.  Computing the integral, this yields the result
\begin{equation}
\mathbb{J}_{\pm}(t)=	\frac{1}{2\pi}e^{-p_0^2-x_0^2} \left(\mp\text{sgn}(\sin\omega t)+\sqrt{\pi } e^{g(x_0, p_0, t)^2} g(x_0, p_0, t) (1\mp\text{sgn}(\sin\omega t)\erf (g(x_0, p_0, t)))\right),\end{equation}
with $g(x_0, p_0, t)=p_0 \cos\omega t -x_0\sin\omega t$.
\subsection{Time derivative of the quasi-probability}
\label{app:curr}
\noindent
To calculate the time-derivative of the QP as appears in Eq.~(\ref{eq:dqdt}), we begin by writing the simple projector identity
\begin{equation}
P\rho + \rho P = P\rho P - \bar P \rho \bar P +\rho,	
\end{equation}
where $\bar P = 1-P$.  Hence the quasi-probability, Eq.~(\ref{eq:qp}) is given by
\begin{equation}
q(-,+)=\frac12\Tr \left(P_{+}(t_2)\left(P_-(t_1) \rho P_-(t_1) - P_+(t_1) \rho P_+(t_1) +\rho\right) \right).
\end{equation}
This may be written in terms of the non-negative sequential measurement probabilities
\begin{equation}
\label{eq:seqprob}
p_{12}(s_1, s_2)=\Tr\left(P_{s_2}(t_2)P_{s_1}(t_1)\rho P_{s_1}(t_1)\right),
\end{equation}
in the form
\begin{equation}
q(-,+)=\frac12\left( p_{12}(-,+)-p_{12}(+,+)+\ev{P_+(t_2)}\right).
\end{equation}
It is now simple to take the derivative with respect to $t_2$, noting that 
\begin{equation}
\frac{d}{dt}\theta(\hat x(t))=\frac{1}{2m}\left(\hat p(t) \delta(\hat x(t)) + \delta(\hat x(t))\hat p(t)\right)=\hat J(t),
\end{equation}
yielding
\begin{equation}
\frac{d q(-,+)}{dt}=\frac12\Tr\left(\hat J(t)\left(P_-(t_1) \rho P_-(t_1) - P_+(t_1) \rho P_+(t_1) +\rho\right) \right).
\end{equation}
We can hence rewrite $q(-,+)$ as
\begin{equation}
\label{eq:qpJ}
q(-,+)=\frac12 \int_{t_1}^{t_2}\mathop{dt} \left(J_-(t)-J_+(t)+J(t)\right),
\end{equation}
where we have introduced the `chopped current' 
\begin{equation}
\label{eq:chopcurr}
J_{\pm}(t)=\ev{\theta(\pm \hat x) \hat J(t) \theta(\pm \hat x)}{\psi},
\end{equation}
which corresponds to the current at the origin, after the initial measurement.  The chopped currents are therefore the currents of the wave-functions $\langle x\lvert e^{-iHt}\theta(\pm\hat x)\rvert \psi\rangle$, and note the connection to the Moshinsky function when $|\psi\rangle$ is expanded in the momentum basis.

Using Eq.~(\ref{eq:class}), we may rewrite Eq.~(\ref{eq:qpJ}) in terms of the difference between quantum and classical post measurement currents, as 
\begin{equation}
q(-,+)=\int_{t_1}^{t_2}\mathop{dt}J_-(t)+\frac12 \int_{t_1}^{t_2}\mathop{dt}\left(\mathbb{J}_-(t)-J_-(t) +\mathbb{J}_+(t)-J_+(t)\right).
\end{equation}

\subsection{Chopped currents calculation}
\label{app:chopcurr}
We calculate the `chopped current' first, that is
\begin{equation}
J_{\pm}(x,t)=\frac{1}{2m}\ev{\delta(\hat x - x)\hat p + \hat p \delta(\hat x - x)}{\phi^{\pm}_\alpha(t)},
\end{equation}
where $\phi^{\pm}_\alpha(t)$ is the time evolution of a coherent state, initially projected on $\theta(\pm \hat x)$ at $t_0$,
\begin{equation}
	\ket{\phi^{\pm}_\alpha(t)} = e^{-iHt}\theta(\pm \hat x)\ket\alpha
\end{equation}
Calculating the current in the position basis, we have
\begin{equation}
J_\pm(x,t)=-\frac{i\hbar}{2m}\left(\phi_\alpha^{\pm*}(x,t)\frac{\partial \phi_\alpha^{\pm}(x,t)}{\partial x}-\frac{\partial \phi_\alpha^{\pm*}(x,t)}{\partial x}\phi_\alpha^{\pm}(x,t)\right),
\end{equation}
equivalent to
\begin{equation}
\label{eqn:Jx}
J_{\pm}(x,t)=\frac{\hbar}{m}\Im\left(\phi_\alpha^{\pm*}(x,t)\frac{\partial \phi_\alpha^{\pm}(x,t)}{\partial x}\right).
\end{equation}
We calculate the evolved chopped state by
\begin{equation}
\label{eq:phipm}
\phi_\alpha^\pm(x,t)=\int_{\Delta(\pm)}\mathop{dy}K(x,y,t)\psi_\alpha(y,t_0),
\end{equation}
where $\Delta(+)=[0,\infty)$, $\Delta(-)=(-\infty,0]$,
\begin{equation}
\psi^{\alpha}(x,t_0)=\frac{1}{\pi^\frac14}\exp(-\frac12(x-x_0)^2 + i p_0 x).
\end{equation}
%\begin{equation}
%\psi_\alpha(x,t)=\left(\frac{m\omega}{\pi\hbar}\right)^\frac14 \exp\left(-\frac{m\omega}{2\hbar}\left(x-\sqrt{\frac{2\hbar}{m\omega}}\Re\alpha(t)\right)^2 + i \sqrt{\frac{2m\omega}{\hbar}}\Im \alpha(t) x\right),
%\end{equation}
is the non-dimenionalized coherent state wave-function, with time evolution $\alpha(t) = e^{-i\omega t}\alpha(0)$, and
\begin{equation}
K(x,y,t)=\left(\frac{1}{2\pi i\sin \omega t}\right)^\frac12 \exp\left(-\frac{1}{2i\sin\omega t}\left((x^2+y^2)\cos\omega t -2 x y\right)\right),
\end{equation}
is the propagator for the harmonic potential.

Inserting the relevant expressions within Eq.~(\ref{eq:phipm}) yields
\begin{multline}
\label{eq:propchop}
\phi_\alpha^\pm(x,t)=\left(\frac{1}{2\pi i \sin \omega t}\right)^\frac12 \left(\frac{1}{\pi}\right)^\frac14 \int_{\Delta(\pm)}\mathop{dy}\exp\left(-\frac12(y-x_0)^2+ i y p_0)\right)\times\\\exp\left(-\frac{x^2 + y^2}{2i \tan \omega t} +\frac{x y}{i \sin \omega t}\right).
\end{multline}
We proceed writing the integral as
\begin{equation}
I_\pm(a,b,c)=\int_{\Delta(\pm)}\mathop{dr}\exp\left(-\frac12(r-a)^2 + i b r+ i c r^2\right),
\end{equation}
where 
\begin{align}
a&=x_0,\\
b&=p_0 - \frac{x}{\sin\omega t},\\
c&=\frac{1}{2\tan\omega t}.
\end{align}
Completing the integration, we have
\begin{equation}
\label{eq:Ipm}
I_\pm(a,b,c)=\frac{\sqrt{\pi } e^{-\frac{2 a^2 c+2 a b+i b^2}{4 c+2 i}} \left(1\pm \erf\left(\frac{a+i b}{\sqrt{2-4 i c}}\right)\right)}{\sqrt{2-4 i c}}.
\end{equation}
We hence can write the chopped wave-function as
\begin{equation}
\label{eq:chopwav}
\phi_\alpha^\pm(x,t)=\left(\frac{1}{2\pi i \sin \omega t}\right)^\frac12 \left(\frac{1}{\pi}\right)^\frac14 e^{-\frac{x^2}{2i \tan \omega t}}I_\pm\left(x_0, p_0-\frac{x}{\sin\omega t},\frac{1}{2\tan\omega t}\right).
\end{equation}
Putting Eq.(\ref{eqn:Jx}) into rescaled units as well, we have
\begin{equation}
\label{eq:Jpmap}
J_{\pm}(x,t)=\omega\Im\left(\phi_\alpha^{\pm*}(x,t)\frac{\partial \phi_\alpha^{\pm}(x,t)}{\partial x}\right).
\end{equation}
To take the derivative we note $I_\pm(a,b,c)$ depends on $s$ only in its second argument, and so we define
\begin{equation}
K_{\pm}(a,b,c)=\frac{\partial b}{\partial s}\frac{\partial}{\partial b}I_\pm(a,b,c),
\end{equation}
which explicitly yields
\begin{equation}
\label{eq:kpm}
K_{\pm}(a,b,c)=\frac{-1}{\sin\omega t}\frac{e^{-\frac{a^2}{2}} \left(-\sqrt{2 \pi } (a+i b) e^{\frac{(a+i b)^2}{2-4 i c}} \left(1\pm \text{erf}\left(\frac{a+i b}{\sqrt{2-4 i c}}\right)\right)\mp 2 \sqrt{1-2 i c}\right)}{2 \sqrt{1-2 i c} (2 c+i)}
\end{equation}

Altogether, this yields the current
\begin{multline}
\label{eq:chopJ}
J_{\pm}(x,t)=\frac{\omega}{2\pi^{\frac32}}\Im\Bigg(\frac{1}{\lvert\sin\omega t\rvert}I_\pm\bigg(x_0,-p_0+\frac{x}{\sin \omega t},-\frac{1}{2\tan \omega t}\bigg)\times\\
\bigg(\frac{ix}{\tan \omega t}I_\pm\bigg(x_0,p_0-\frac{x}{\sin \omega t},\frac{1}{2\tan \omega t}\bigg) +K_\pm\bigg(x_0,p_0-\frac{x}{\sin \omega t},\frac{1}{2\tan \omega t}\bigg)\bigg)
\Bigg).
\end{multline}
We also calculate the current of the original unperturbed coherent state, in these same rescaled units.  Since coherent states are eigenfunctions of the annihilation operator, with eigenvalue $\alpha(t)$, it follows that 
\begin{equation}
\frac{\partial}{\partial x} \psi_\alpha(x,t)=\psi_\alpha(x,t)\left(\sqrt2 \alpha(t) - x\right).
\end{equation}
Hence the current is given by 
\begin{equation}
J(x,t)=\frac{\omega}{\sqrt\pi}\Im\left(e^{-(x-\sqrt{2}\Re \alpha(t))^2}(\sqrt2 \alpha(t) -x)\right),
\end{equation}
where taking the imaginary part, and using Eq.~(\ref{eq:alphatox}), we have
\begin{equation}
	J(x,t)=\frac{\omega}{\sqrt\pi}p_t e^{-(x-x_t)^2}.
\end{equation}
\section{Small Time Current Expansions}
\label{app:tJ}
We are interested in a small time expansion of the chopped current $J_{\pm}(t)$, for any general state $\ket\psi$.  We start by defining the chopped-evolved state,
\begin{equation}
	\ket{\psi^{\pm}(t)}=e^{-iHt}\theta(\pm \hat x)\ket{\psi}.
\end{equation}
 We now follow Appendix~\ref{app:curr} up to Eq.~(\ref{eq:Jpmap}) to calculate the current, however this time using the general $\ket\psi$ state, leading to
\begin{equation}
J_{\pm}(t)=\omega\Im\left(\psi^{\pm*}(0,t)\frac{\partial\psi^\pm(0,t)}{\partial x}\right),
\end{equation}
with $\ket\psi$ represented in the non-dimensional position basis, i.e. $\braket{x}{\psi}=\left(\frac{m\omega}{\hbar}\right)^{\frac{1}{4}}\psi(x)$, with a normalised $\psi(x)$ which is purely a function of $x$.

Using the QHO propagator as in App.~\ref{app:curr} up to Eq.~(\ref{eq:propchop}), now with the $\ket\psi$ state, we have
\begin{equation}
\label{eq:genprop}
\psi^{\pm}(x,t)=\left(\frac{1}{2\pi i \sin\omega t}\right)^\frac12 \int_{\Delta(\pm)}\mathop{dy}\psi(y,0)\exp\left(-\frac{x^2 + y^2}{2i \tan \omega t} +\frac{x y}{i \sin \omega t}\right).
\end{equation}
We interested in the current at $x=0$ which means we only need $\psi(x)$ and its first derivative evaluated at $x=0$, leaving
\begin{equation}
\label{eq:psi0}
	\psi^{\pm}(0,t)=\left(\frac{1}{2\pi i \sin\omega t}\right)^\frac12 \int_{\Delta(\pm)}\mathop{dy}\psi(y,0)\exp\left(-\frac{y^2}{2i \tan \omega t}\right).
\end{equation}
We now argue that for small $t$ the main contribution to the integral will come from near the boundary of the chop,  and hence Taylor expand  $\psi(y,0)$ expand around $y=0$.  By using the parametrisation $y=z(\tan\omega t)^\frac12$, we simplify the exponential part of the integrand.
\begin{equation}
\psi(y,0)=\sum_{n=0}^{\infty}\frac{\psi^{(n)}(0,0)}{n!}z^n(\tan\omega t)^\frac{n}{2}.
\end{equation}
Using this in Eq.~(\ref{eq:psi0}), and using the substitution within the integral, we have
\begin{equation}
	\psi^{\pm}(0,t)=\left(\frac{\tan\omega t}{2\pi i \sin\omega t}\right)^\frac12 \int_{\Delta(\pm)}\mathop{dz}\exp\left(-\frac{z^2}{2i}\right)\sum_{n=0}^{\infty}\frac{\psi^{(n)}(0,0)}{n!}z^n(\tan\omega t)^\frac{n}{2}
\end{equation}
Interchanging the order of summation and integration, we have
\begin{equation}
\psi^\pm(0,t)=\left(\frac{1}{2\pi i \cos\omega t}\right)^\frac12\sum_{n=0}^\infty \frac{\psi^{(n)}(0,0)}{n!}(\tan \omega t)^\frac{n}{2}\int_{\Delta(\pm)}\mathop{dz}e^{-\frac{z^2}{2i}}z^n.
\end{equation}
We now define
\begin{equation}
K_\pm(n)=\int_{\Delta(\pm)}\mathop{dz}e^{-\frac{z^2}{2i}}z^n,
\end{equation}
which can be calculated by taking the Fourier transform of $e^{iz^2}\theta(\pm z)$, taking $n$ derivatives in Fourier space, and then evaluating with the conjugate variable set to $0$, to give the result
\begin{equation}
K_{\pm}(n)=(\pm1)^n 2^{\frac{n-1}{2}} e^{\frac{1}{4} i \pi  (n+1)} \Gamma \left(\frac{n+1}{2}\right),
\end{equation}
were $\Gamma$ is the Gamma function. This gives a final result of
\begin{equation}
\label{eq:psiexp0}
\psi^\pm(0,t)=\left(\frac{1}{2\pi i \cos \omega t}\right)^\frac12 \sum_{n=0}^\infty K_\pm(n)\frac{\psi^{(n)}(0,0)}{n!}(\tan\omega t)^\frac{n}{2}.
\end{equation}
We now calculate the derivative of the chopped wavefunction, by taking the derivative of Eq.~(\ref{eq:genprop}), evaluated at $x=0$ to find
\begin{equation}
\frac{\partial \psi^\pm(x,t)}{\partial x}\eval_{x=0}	=\frac{-i}{\sin \omega t}\left(\frac{1}{2\pi i \sin\omega t}\right)^\frac12 \int_{\Delta(\pm)}\mathop{dy}\psi(y,0)y \exp\left(-\frac{y^2}{2i \tan \omega t}\right).
\end{equation}
We now note, that with the exception of the pre-factor $\frac{-i}{\sin\omega t}$, this is the same result as before, only with $\psi(y,0)$ swapped for $y\psi(y,0)$, leading to the result
\begin{equation}
\frac{\partial \psi^\pm(0,t)}{\partial x}=-\left(\frac{-1}{2\pi i\sin^2\omega t \cos\omega t}\right)^\frac12\sum_{n=0}^\infty K_\pm(n)\frac{\frac{\partial^n}{\partial y^n}(y\psi(y,0))\eval_{y=0}}{n!}(\tan\omega t)^\frac{n}{2}.
\end{equation}
Then since 
\begin{equation}
\frac{\partial^n}{\partial y^n}y\psi(y,0)\eval_{y=0}=n \psi^{(n-1)}(0,0),
\end{equation}
we have as a final result 
\begin{equation}
\label{eq:psipexp0}
\frac{\partial \psi^\pm(0,t)}{\partial x}=-\left(\frac{-1}{2\pi i\sin^2\omega t \cos\omega t}\right)^\frac12\sum_{n=1}^\infty K_\pm(n)\frac{n\psi^{(n-1)}(0,0)}{n!}(\tan\omega t)^\frac{n}{2},
\end{equation}	
noting the change on the sum's lower limit.

We now combine Eq.~(\ref{eq:psiexp0}) and Eq.~(\ref{eq:psipexp0}) to yield the small-time expansion for the chopped current
\begin{equation}
J_\pm(t)=\frac{-\omega}{2\pi\sin\omega t \cos\omega t}\Im \left(i\sum_{\ell=0}^{\infty}\sum_{n=1}^{\infty}K_{\pm}(n)K^*_{\pm}(\ell) \frac{n\psi^{(n-1)}(0,0)\psi^{*(\ell)}(0,0)}{n!\ell!}(\tan\omega t)^\frac{n+\ell}{2}\right)
\end{equation}
We note that this result is in fact trivial to integrate over time by noting the derivative of $\tan(\omega t)$, yielding
\begin{equation}
\int_{0}^{\tau}\mathop{dt}\frac{(\tan\omega t)^\frac{k}{2}}{\sin\omega t\cos\omega t}=\frac{2}{k\omega}(\tan\omega \tau)^{\frac{k}{2}}.
\end{equation}
The time-integral of the chopped current is thus
\begin{equation}
\int_{0}^{\omega \tau}J_\pm(t)\mathop{dt}=-\frac{1}{\pi}\Im\left(i\sum_{\ell=0}^{\infty}\sum_{n=1}^{\infty}K_{\pm}(n)K^*_{\pm}(\ell) \frac{n\psi^{(n-1)}(0,0)\psi^{*(\ell)}(0,0)}{(n+\ell)n!\ell!}(\tan\omega t)^\frac{n+\ell}{2}\right)
\end{equation}
We also note that by defining
\begin{equation}
L(n)=K_+(n)+K_-(n),	
\end{equation}
we may adapt the result to the unchopped current as
\begin{equation}
J(t)=-\frac{\omega}{2\pi\sin\omega t \cos\omega t}\Im \left(i\sum_{\ell=0}^{\infty}\sum_{n=1}^{\infty}L(n)L^*(\ell) \frac{n\psi^{(n-1)}(0,0)\psi^{*(\ell)}(0,0)}{n!\ell!}(\tan\omega t)^\frac{n+\ell}{2}\right),
\end{equation}
with time-integral
\begin{equation}
\int_{0}^{\omega \tau}J(t)\mathop{dt}=-\frac{1}{\pi}\Im\left(i\sum_{\ell=0}^{\infty}\sum_{n=1}^{\infty}L(n)L^*(\ell) \frac{n\psi^{(n-1)}(0,0)\psi^{*(\ell)}(0,0)}{(n+\ell)n!\ell!}(\tan\omega t)^\frac{n+\ell}{2}\right)
\end{equation}
Using Eq.~(\ref{eq:qpJ}) to express the quasi-probability as the time integral of currents, and noting the similarity of the summands, we can write the quasi-probability for as
\begin{equation}
q(-,+)=-\frac{1}{2\pi}\Im\left(i\sum_{\ell=0}^{\infty}\sum_{n=1}^{\infty}\mathcal{Q}(n,\ell) \frac{n\psi^{(n-1)}(0,0)\psi^{*(\ell)}(0,0)}{(n+\ell)n!\ell!}(\tan\omega t)^\frac{n+\ell}{2}\right),
\end{equation}
where
\begin{equation}
\mathcal{Q}(n,\ell)=K_-(n)K_-^*(\ell)-K_+(n)K_+^*(\ell)+L_n L^*_\ell.
\end{equation}
By approximating the infinite sums to finite order, we are able to approximate the quasi-probability.  To get the first three terms of the approximation, we limit both sums to $\ell_{max}=2$ and $n_{max}=3$, yielding
\begin{multline}
\label{eq:tayl3}
q(-,+)=	\frac{1}{2\sqrt{\pi}}\lvert\psi(0,0)\rvert^2 \tan^\frac12\omega\tau + \frac{J(0)}{2}\tan\omega\tau+\\\frac{1}{6\sqrt\pi}\left(\lvert \psi'(0,0)\rvert^2-\left[\frac{1}{4}+\frac{3 i}{4}\right]\psi''^{*}(0,0)\psi(0,0)- \left[\frac{1}{4}-\frac{3 i}{4}\right]\psi''(0,0)\psi^*(0,0)\right)\tan^\frac32\omega t \\+\mathcal{O}(\tan ^2 \omega t)
\end{multline}
Taking the taking the $\omega\rightarrow 0$ expansions of the trigonometric terms recovers the result for the free particle,
\begin{multline}
q(-,+)=\frac{1}{2\sqrt{\pi}}\lvert\psi(0,0)\rvert^2 \tau^\frac12+ \frac{J(0)}{2}\tau+\\\frac{1}{6\sqrt\pi}\left(\lvert \psi'(0,0)\rvert^2-\left[\frac{1}{4}+\frac{3 i}{4}\right]\psi''^{*}(0,0)\psi(0,0)- \left[\frac{1}{4}-\frac{3 i}{4}\right]\psi''(0,0)\psi^*(0,0)\right)\tau^\frac32,
\end{multline}
which we note has a term in $\tau^\frac12$ which was missing from an earlier calculation of this expansion in Ref.~\cite{halliwell2019c}, as well as a different coefficient on the $\tau^\frac32$ term.

\begin{figure}
\includegraphics[height=5.3cm]{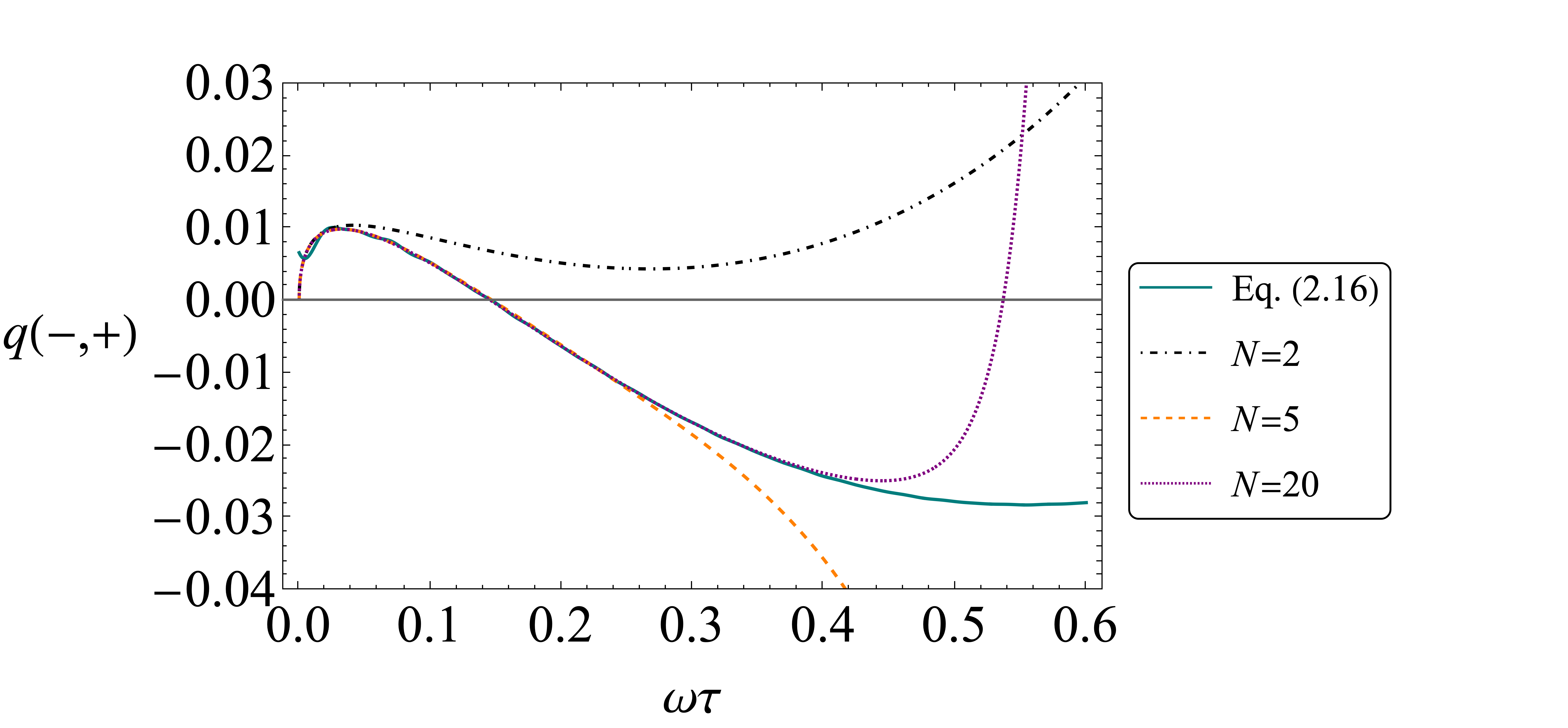}
	\caption{The small-time expansion of $q(-,+)$ plotted alongside the previous calculation, at varying degrees of truncation, where $N=2$ corresponds to Eq.~(\ref{eq:tayl3}).}%%
	\label{fig:tayl}
\end{figure}

The initial divergence of the quantum chopped current is clearly seen. These results also agree with the small time expansion of chopped currents given by Sokolowski \cite{sokolovski2012}, giving another useful check on our calculations. For our gaussian initial state we find agreement with the results above, we plot this expansion alongside our original calculation in Fig.~\ref{fig:tayl}.

\section{Wigner Function calculational details}
\label{app:wig}
The Wigner-Weyl transform, which maps Hermitian operators to real phase space functions~\cite{hillery1984a, tatarskii1983a, case2008, halliwell1993a}, is defined by
\begin{equation}
\label{eq:wigtx}
	W_A(X,p) = \frac{1}{2\pi}\int_{-\infty}^{\infty}\mathop{d\xi}e^{-ip\xi}\mel{X+\tfrac{\xi}{2}}{A}{X-\tfrac{\xi}{2}}.
\end{equation}
Traces of pairs of operators may be expressed in the Wigner representation as
\begin{equation}
\label{eq:wigexp}
	\Tr(\hat A\hat B) = 2\pi \int_{-\infty}^{\infty}\mathop{dX}\int_{-\infty}^{\infty}\mathop{dp}W_{A}(X,p)W_{B}(X,p).
\end{equation}
To apply this formula to the quasi-probability there are two natural ways to proceed. First, in Ref.~\cite{halliwell2019c}, the free particle quasi-probability was explored in Wigner-Weyl form using $\hat A = \tfrac12\left(P_{s_1}P_{s_2}+P_{s_2}P_{s_1}\right)$ and $\hat B=\rho$. However this was not found to be very useful since $W_A(X,p)$ in this case is highly oscillatory and it was not possible to clearly identify the regions of negativity, hence we proceed with a different approach.

We first write the QP in the form $q(s_1, s_2)=\Tr(\bar{\rho}_{s_1}P_{s_2}(\tau))$, where 
 $\bar{\rho}_{s_1}=\frac12(P_{s_1}\rho+\rho P_{s_1})$, and $t_1=0$ without loss of generality.    
Hence by Eq.~(\ref{eq:wigexp}) we have 
\begin{equation}
\label{eq:wigqp}
	q(s_1,s_2)=2\pi \int_{-\infty}^{\infty}\mathop{dX}\int_{-\infty}^{\infty}W_{\bar\rho_{s_1}}(X,p)W_{P_{s_2}(\tau)}(X,p),
\end{equation}
where $W_{P_{s_2}}(X,p)=\theta(s_2(X \cos \omega t + p \sin\omega t))$.  Using Eq.~(\ref{eq:wigtx}), the transform of $\bar\rho_{s_1}$ is given by
\begin{equation}
\label{eq:wigrhob}
W_{\bar\rho_{s_1}}(X,p)=\frac{1}{4\pi}\int_{-\infty}^{\infty}\mathop{d\xi}e^{-ip\xi}\mel{X+\tfrac{\xi}{2}}{P_{s_1}\rho+\rho P_{s_1}}{X-\tfrac{\xi}{2}},
\end{equation}
We without loss of generality we take $t_1=0$, and so $P_{s_1}=\theta(s_1 \hat x)$, leading to
\begin{equation}
W_{\bar\rho_{s_1}}(X,p)=\frac{1}{4\pi}\int_{-\infty}^{\infty}\mathop{d\xi}e^{-ip\xi}\left(\theta(s_1(X+\tfrac{\xi}{2}))+\theta(s_1(X-\tfrac{\xi}{2}))\right)\mel{X+\tfrac{\xi}{2}}{\rho}{X-\tfrac{\xi}{2}}.
\end{equation}
Since coherent states are pure states, we have simply that $\rho(x,y)=\psi(x)\psi^*(y)$, where we will use natural units with $\psi(x)=\tfrac{1}{\pi^\frac14}\exp(-\tfrac12(x-x_0)^2+ip_0x)$.  This yields the integrand as $I(X,\xi)=e^{-ip\xi}\exp\left(i p_0 \xi-X^2+2 X x_0-x_0^2-\tfrac{\xi^2}{4}\right)$.  

Demonstrating with the $s_1=+1$ case, the theta functions are handled by splitting the integral into two integrals over the regions $[-2X, \infty)$, and $(-\infty, 2X]$, and so we have
\begin{equation}
W_{\bar\rho_{s_1}}(X,p)=\frac{1}{4\pi\sqrt{\pi}}\left(\int_{-2X}^{\infty}\mathop{d\xi}I(X,\xi)+\int_{-\infty}^{2X}\mathop{d\xi}I(X,\xi)\right),
\end{equation}
Computing the integral involved, we reach the result
\begin{equation}
W_{\bar\rho_{s_1}}(X,p)=\frac{1}{2\pi}e^{-(p-p_0)^2- (X-x_0)^2}\left(1+\Re\erf\left(i(p-p_0)+s_1 X\right)\right),
\end{equation}
which may be written as
\begin{equation}
\label{eq:wigint}
W_{\bar\rho_{s_1}}(X,p)=\frac12 W_{\rho}(X,p)\left(1+\Re\erf\left(i(p-p_0)+s_1 X\right)\right).
\end{equation} 
in terms of $W_{\rho}(X,p)$, the Wigner function of the pure coherent state, given by
\begin{equation}
	W_{\rho}(X,p)=\frac{1}{\pi}\exp(-(p-p_0)^2-(X-x_0)^2).
\end{equation}
The classical equivalent for Eq.~(\ref{eq:wigint}) is $\frac12 W_\rho(X,p)(1+\text{sgn}(X))$, which is approached for $p$ close to $p_0$ and for large $\lvert X \rvert$.

The time evolution of the Wigner function in the case of the QHO is given by rigid rotation in accordance with classical paths $X_{\tau}=x_0\cos \omega \tau - p_0 \sin\omega\tau$.  Hence in Eq.~(\ref{eq:wigqp}) $W_{P_{s_2}(\tau)}(X,p)=\theta(s_2 X_{-\tau})$, and the final expression for the quasi-probability is
\begin{equation}
	q(s_1,s_2)=\int_{-\infty}^{\infty}\mathop{dX}\int_{-\infty}^{\infty}\mathop{dp}f_{s_1, s_2}(X,p)
\end{equation}
with the phase-space density $f_{s_1, s_2}(X,p)$ given by
\begin{equation}
\label{eq:wigfin}
f_{s_1, s_2}(X,p)=\frac12 W_{\rho}(X,p)\left(1+\Re\erf\left(i(p-p_0)+s_1 X\right)\right)\theta(s_2 X_{-\tau}).
\end{equation}

\begin{figure}[t]
	\subfloat[]{{\includegraphics[width=7.6cm,trim={0 1.5cm 3cm 4cm},clip]{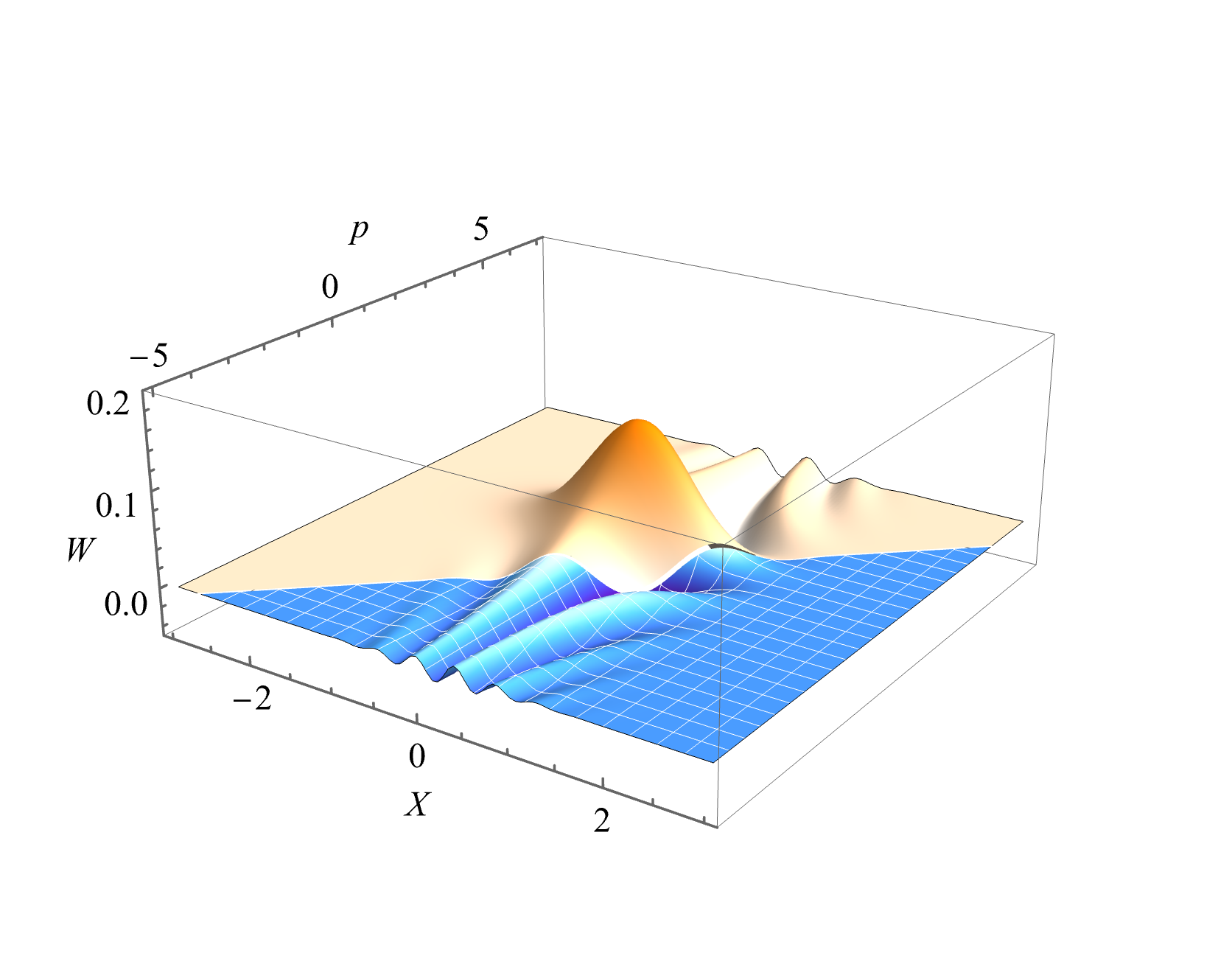}}}%
	%\qquad
	\hspace{5mm}
	\subfloat[]{{\includegraphics[width=7.6cm,trim={3.9cm 0 3.9cm 0},clip]{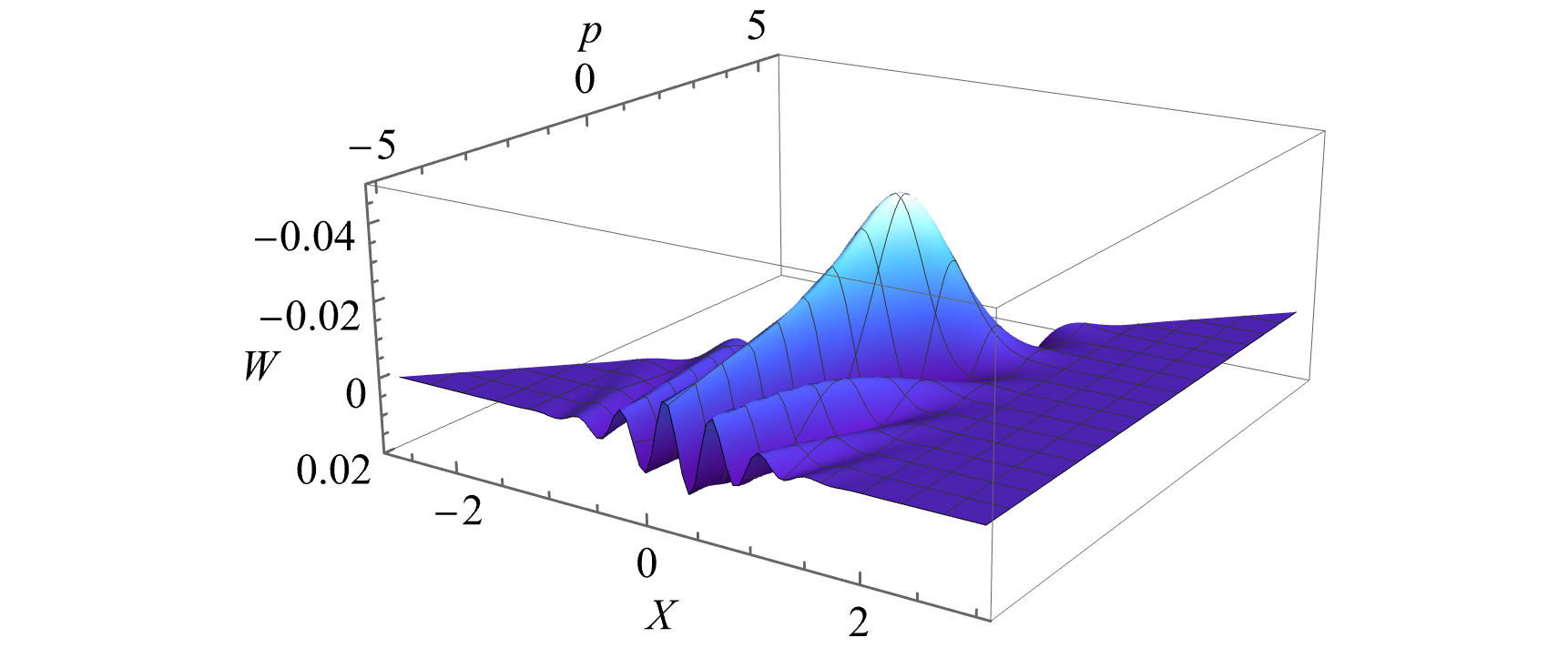}}}
	\caption{Plot (a) shows $W_{\bar\rho_{-}}(X,p)$, Eq.(\ref{eq:wigint}) with $s_1=-1$, for an initial coherent state with $x_0=0.55$, $p_0=-1.925$ and $\omega\tau=0.55$. The orange (smooth) region shows the region removed when we go to the phase space density Eq.(\ref{eq:wigfin}), thereby showing how the most significant positive parts are removed. Plot (b) (flipped axis) shows the phase space density Eq.(\ref{eq:wigfin}) which will from visual inspection integrate to a negative number overall.}%
	\label{fig:wig2}%
\end{figure}
Eq. (\ref{eq:wigint}) and Eq. (\ref{eq:wigfin}) are plotted in Fig.~\ref{fig:wig2}.

\bibliography{paper}

\end{document}